\documentclass[twocolumn]{emulateapj}
\usepackage{amssymb}
\usepackage{amsmath}
\usepackage{graphicx}
\usepackage{natbib}
\usepackage{bigints}
\usepackage{relsize}
\usepackage{multirow}
\usepackage{color}
\usepackage[flushleft]{threeparttable}

\newcommand{\myemail}{radigan@stsci.edu}

\shorttitle{Independent analysis of BAM survey data}
\shortauthors{J. Radigan}

\begin{document}


\title{An independent analysis of the brown dwarf atmosphere monitoring (BAM) data: large-amplitude variability is rare outside the L/T transition}
\author{Jacqueline Radigan\altaffilmark{1}}

\altaffiltext{1}{Space Telescope Science Institute, 3700 San Martin Drive Baltimore MD 21218; \myemail}

\begin{abstract}
Observations of variability can provide valuable information about the processes of cloud formation and dissipation in brown dwarf atmospheres.  Here we report the results of an independent analysis of archival data from the Brown dwarf Atmosphere Monitoring (BAM) program.  Time series data for 14 L and T dwarfs reported to be significantly variable over timescales of hours were analyzed. We confirm large-amplitude variability (amplitudes $>$2\%) for 4/13 targets and  place upper limits of 0.7\%-1.6\% on variability in the remaining sample.  For two targets we find evidence of weak variability at amplitudes of 1.3\% and 1.6\%.  Based on our revised classification of variable objects in the BAM study, we find strong variability outside the L/T transition to be rare at near infrared wavelengths.  From a combined sample of 81 L0-T9 dwarfs from the revised BAM sample and the variability survey of Radigan et al. we infer an overall observed frequency for large-amplitude variability outside the L/T transition of 3.2$_{-1.8}^{+2.8}$\%,  in contrast to 24$^{+11}_{-9}$\% for L9-T3.5 spectral types.  We conclude that while strong variability is not limited to the L/T transition, it occurs more frequently in this spectral type range, indicative of larger or more highly contrasting cloud features at these spectral types.
\end{abstract}

 \section{Introduction}

Time domain observations of brown dwarfs provide a method for probing cloud structure in cool atmospheres\citep{tinney99,bailer-jones02,artigau09,radigan12,buenzli12,apai13,heinze13,biller13,burgasser14}.  Statistical properties of variability as a function of spectral type can in turn shed light on mechanisms responsible for the formation, evolution, and dissipation of condensate clouds across the M-L-T-Y sequence.  A key question that has motivated much of the work to date is whether variability may be more common at the transition between cloudy L, and clear T spectral types; the development of cloud holes has been put forward as one possible explanation for the abrupt decline in condensate opacity and NIR color reversal from late-L to mid-T spectral types \citep{ackerman01,burgasser02_lt,marley10}.  A testable consequence of this hypothesis is that brown dwarfs occupying the L/T transition color reversal may display greater levels of variability than other brown dwarfs.

An overview of previous monitoring campaigns for brown dwarf variability can be found in \citet{radigan14}, to which the reader is referred for more detailed background.  At near-infrared (NIR) wavelengths, which are sensitive to changes in cloud opacity, photometric monitoring of brown dwarfs has revealed low-level variability to be common across the entire L-T sequence \citep{koen04a,koen05a,clarke08,radigan14}.  Extrapolations from space-based surveys suggest that most, if not all, brown dwarfs have patchy photospheres to some degree \citep[][Metchev et al. 2014, submitted]{buenzli14}.

High-precision surveys from the ground have shown that when variability is detected, amplitudes are typically $<$2\% in the $J$-band \citep[based on 31 targets from][]{koen04a,koen05a,clarke08,girardin13}.  An exception to this is a handful of early T-dwarfs that have been reported to vary with amplitudes of several percent (and in one case, up to 26\%): 2M2139$+$02 (T1.5), SIMP~0136$+$09 (T2.5), Luhman 16B (T0.5), SDSS1052$+$21 (T1/T1), SIMP1629$+$03 (T2), and 2M0758$+$32 (T2) \citep{artigau09,radigan12,gillon13,biller13,girardin13,radigan14}. The early T-dwarfs fall in the milieu of the NIR color reversal that characterizes the L/T transition \citep[e.g.][]{dupuy12,faherty12}.  Notably, these spectral types are largely absent in earlier surveys discussed above,\footnote{ \citet{koen04a,koen05a} monitored one early T-dwarf, as well as the T1/T6 binary Eps Indi BC, finding evidence for large changes in the nightly mean flux of the latter in the H-band; \citet{clarke08} monitored the L6/T2 binary 2M0423 and found evidence of variability  with a 0.8\% amplitude, which given the flux ratio of $\Delta m$=0.4 for the system would equate to either a 1.4\% variation of the L6 component or a 2.0\% variation of the T2 component; \citet{girardin13} monitored 4 L9-T3.5 dwarfs and found the T1/T1 binary SDSS 1052$+$21 to be a large-amplitude variable.}  perhaps explaining why a clear-cut example large-amplitude variability was not published until 2009, and came from a targeted observation of a single object \citep{artigau09}.  

Two additional studies by \citet{enoch03} and \citet{khandrika13}, monitored 9 and 11 L and T dwarfs respectively in the $J$ and $K_s$ bands.  Both surveys reported $\sim$1/3 of targets to be variable with amplitudes $>$10\% at levels of $\sim$2-3 times the formal photometric uncertainties.  However, these findings have not been supported by earlier or subsequent high-precision investigations. 

While low-amplitude NIR variability appears to be common for brown dwarfs, the present analysis focusses on large-amplitude signals, defined here to mean amplitudes $>$2\%.  The work presented here asks the following questions: (1)what is the frequency of large-amplitude variability for brown dwarfs, and (2)how are large-amplitude variables distributed as a function of spectral type?  With the recent publication of two large, high-precision, variability surveys in the NIR by \citet{radigan14} (R14) and \citet{wilson14} (W14 hereafter), for which large-amplitude signals are broadly detectable, we are now well-poised to investigate these questions.  However, the studies of R14 and W14, which are similar in terms of photometric precision and observing timescale, appear to reach different conclusions. 
 
 The study of R14 surveyed 62 L4-T9 dwarfs in uninterrupted sequences of $\sim$2-5\,hr using wide field infrared cameras on the 2.5-m Dupont telescope, and the Canada-France-Hawaii Telescope.   Large-amplitude variability was found to be more prevalent among early T-dwarfs.   Of 57 objects included in the statistical sample, large-amplitude variability was reported for 4/16 objects within the L/T transition color reversal (defined in R14 as L9-T3.5 spectral types), in comparison 0/41 objects at all other spectral types.  Outside the L/T transition, R14 report significant variability in 5 additional targets (4 mid-T dwarfs and a mid-L dwarf) at lower amplitudes of $0.6-1.6$\%, inferring low-level variability to be common for brown dwarfs at all spectral types.

The Brown dwarf Atmosphere Monitoring (BAM) program of W14 used the SofI instrument on the 3.5-m New Technology Telescope (NTT) to survey 69 L0-T9 dwarfs over timescales of $\sim$2-4\,hr, switching back and fourth between two targets at a time, every $\sim$15 min.   Significant variability is reported for 14 targets spanning the entire early-L to late-T spectral sequence with peak-to-trough amplitudes ranging from 1.7\%-11\% (13 of which have amplitudes exceeding 2\%, and therefore meet the R14 criterion for large-amplitude variability). The frequency of variability inside the L/T transition was found to be indistinguishable from the frequency outside this region.

While the overall number of variables reported by R14 and W14 are similar, their distributions of amplitudes are significantly different.  The high number of large-amplitude variables reported by W14 (especially outside the L/T transition) is inconsistent with the findings of R14 and those of earlier high-precision studies.

Here we present  an independent reduction and analysis of the archival time series data for the 14 targets reported to be significantly variable by W14.  Our procedure for reducing the NTT/SofI  data is detailed in section \ref{sect:reduc}.  Based on the resultant light curves presented in section \ref{sect:results} and the Appendix, we find evidence of large-amplitude variability for 4 objects, placing upper limits of $\sim$1\%-1.6\% on the variability of the remaining targets.  In section \ref{sect:discuss}, based on the results of our analysis, as well as null detections reported in W14 we calculate occurrence rates of large-amplitude variability for L, T, and L/T transition dwarfs respectively in the revised BAM sample (rBAM), as well as the combined rBAM$+$R14 sample.

 \label{sect:intro}

\section{Data Reduction}
\label{sect:reduc}
Data for the BAM survey (program 188.C-0493) obtained during two separate observing runs as described in W14 was downloaded directly from the European Space Observatory (ESO) archive using a coordinate query to specify targets of interest.  Data and corresponding calibration files were downloaded for the 14 targets reported to be variable by W14, listed in table \ref{tab:targets}. Calibration files consist of a set of ``special'' dome flats, as well images of a standard star observed on a 4$\times$4 grid used to correct for low-frequency non-flatness of the dome flat fields, due to non-uniform illumination of the screen.  Since the bias of the SofI detector is illumination-dependent the dark$+$bias contribution to each science image is subtracted using a sky frame, rather than a non-illuminated dark frame.  The reduction of all science and calibration images was accomplished using custom $IDL$ routines that will be made available online\footnote{http://www.stsci.edu/$\sim$radigan/sofi/}.

\subsection{Crosstalk Removal}
The SofI detector suffers inter-quadrant crosstalk along rows, such that a bright source in the upper quadrants of the array produces a faint glow in the the equivalent rows of the lower quadrants, and vice versa.  The crosstalk effect scales with the total intensity along a given row in the opposite quadrants, by an empirically determined coefficient $\alpha$.  We have used a value of $\alpha=1.4 \times10^{-5}$ as recommended by the ESO online documentation, and do not observe any residual crosstalk artifacts in the data.

\subsection{Special Dome Flats}
Due to the illumination dependent bias or ``shade'' in SofI images, the subtraction of lamp-off from lamp-on dome flats will result in a residual shade pattern varying by a few percent along columns of the array.  In order to estimate the illumination-dependent shade pattern, a special sequence of dome flats are obtained with the focal plane mask aligned and then misaligned.  The obscured region of the misaligned frames allows the shade pattern to be estimated independently, and removed from the aligned frames before they are combined in the standard way, as described by the ESO online documentation and in \citet{tinney03}.  

\subsection{Illumination Correction}
An illumination correction can be used to correct for low-frequency non-flatness of the special dome flats by observing a standard star in a 16-position grid across the array.  The grid images are sky-subtracted and flattened with the special dome flat, and subsequently the standard star flux is measured at all grid positions.  A 2D surface (2nd order polynomial in our case) is fit to the flux grid to estimate the illumination correction.  The special dome flat can then be corrected via multiplication with the illumination correction surface.  The BAM observations used a compact dither pattern, and in most cases there is no need for low-frequency corrections to the flat field.  For all targets reduction was tested both with and without the illumination correction, and in most cases it was found to have little to no effect.  For a few targets there is a change in mean position over the course of the observation, and in these cases the illumination correction slightly improves the photometric precision (at the sub-percent level).  There is no instance where the application of this correction changes the qualitative nature of the light curves obtained, or induces variability in a previously non-variable target or vice-versa.  The correction has been applied to the special dome flats used to calibrate the data products presented in this paper.

\subsection{Sky Subtraction}
The sky subtraction of SofI images serves to remove the dark current as well as the flux-dependent shade pattern.  Sky frames are created by median combining a set of dithered science images.  In the ideal case all the sky images would have identical illumination, allowing the true sky$+$dark$+$shade to be recovered, while filtering out stars.  However, the sky brightness can change substantially over several minutes, especially at the beginning and end of the night.  In practice the sky frames can be normalized before median combining, and then re-scaled to the science frame, which preserves the structure of the sky, but leaves a residual shade pattern.  If the sky frames are median combined without any scaling and subtracted directly from the science images the shade pattern is more cleanly removed at the expense of the sky removal.  Both strategies were attempted with the BAM data, and the images produced using scaled and median-combined sky frames yielded slightly better photometric precision in the end products.

Sky subtraction was achieved using a running sky frame, where science frames closest in time to (within 15\,min), and sufficiently offset ($>$19-35 pixels) from the frame in question are identified and use to build a sky.  In addition to the time and offset requirements for the running sky frames, we also implemented an illumination requirement (overall differences in illumination of  $<$5\%-15\%) in order to minimize the residual shade pattern where possible.  

\subsection{Pipeline Workflow}
The workflow of the $IDL$ pipeline proceeds in the following manner.  First, the crosstalk correction is applied to all raw science and calibration images.  Next, a first pass sky subtraction is performed on all science images as described above.  The sky-subtracted images are then divided by the illumination-corrected flat field.  Target and reference star positions are then determined via point-and-click user input with subsequent centroiding.  Finally, a second pass sky subtraction$+$flat fielding is performed (discarding the first-pass products) masking out the target and reference stars from the running sky frames.   

\subsection{Aperture Photometry}
Aperture photometry was performed on the target and reference stars identified in the reduced science images using the IDL routine {\tt APER}.  Before performing photometry stellar centroids were measured in each image using both gaussian PSF fitting and the $IDL$ {\tt CNTRD} procedure, which computes where x- and y- derivatives go to zero, based on the DAOPHOT algorithm.  Both methods agree, and the PSF-fitting centroids are used in this analysis.  The final photometry is largely unaffected by this choice.  The use of fixed-size photometry apertures, as well as ones that scale with the median FWHM of all stars on chip were tested, with fixed apertures producing the most stable photometry as gauged by the behavior of reference stars.  \citet{clarke08} also apply a fixed photometry aperture to SofI data with good results.  For each sequence aperture photometry was performed for an array of fixed aperture sizes, and the aperture that minimizes the relative flux RMS of a set of iteratively chosen, well-behaved reference stars, was chosen for the final light curves.  

\subsection{Light Curve Analysis}
High-precision differential photometry is accomplished by identifying a common systematic trend (i.e. due to atmosphere, sky, instrumental changes) using an ensemble of reference stars, and then dividing this trend out of the target light curve, leaving (in the ideal case) only intrinsic astrophysical variability.  In practice not all reference stars are well behaved and conform to the global trend (e.g. due to differences in PSF, falling on a noisy part of the array, background contamination, etc), so care must be taken in choosing a good set of reference stars. Furthermore, even after dividing out the global trend determined from a set of well-behaved references, residual systematics can remain in the corrected target light curves.  Care must therefore be taken to understand the the amplitude of such residual systematic effects.  This can be achieved by cycling through each reference star and correcting it in a similar manner as for the target, using the remaining ensemble of reference stars (less the target and star in question) to determine the global trend.  Then, each corrected reference star acts as a control, from which the level of residual systematics can be inferred.  This is complicated by the fact that the reference stars may themselves be variable, and requires either (i)a comprehensive statistical analysis of a large number of representative reference stars as done by \citet{gelino02,heinze13,koen13,radigan14}, or (ii)less rigorously (applicable for a smaller data set) user discretion, guided by common sense.  For instance, if the bulk of reference stars are flat, but one or two stand out as variable, then astrophysical variations may account for these outliers.  However, the amplitude at which most or many reference stars exhibit variations (which can be a function of source brightness), likely represents a systematic noise floor of the data, and similar variations in the target have no meaningful interpretation.

For each monitoring sequence, correction of the raw flux light curves was performed following a method similar to that described in \citet{radigan12,radigan14}.  First, all raw flux light curves were normalized by their median values.  An estimate of the global trend, which is referred to as the calibration curve hereafter, is constructed by median-combining the normalized raw-flux light curves of an ensemble of reference stars.  An independent calibration curve is constructed for the target as well as each reference star light curve that will serve as a comparison/control.  The target is always excluded from the ensemble of stars used to compute calibration curves, and each reference star is removed from the set of stars used to compute its own correction.  After a first pass correction, properties of the target and reference star light curves are determined, and noisy or variable reference stars are removed from all calibration curves in subsequent passes.   Reference stars whose corrected light curve RMS exceeds $\sim$1.5-2.5$\times$ that of the target light curve are typically excluded, and this number is adjusted upward when required to obtain at minimum two (preferably more) good reference stars.  The above procedure is iterated upon 3 times, which is sufficient to converge upon a stable set of well-behaved reference stars.  A final manual inspection of the corrected reference star light curves is performed, where additional references may be excluded in a final pass if they display large-amplitude trends despite having a light curve RMS below the automated cut-off.

\begin{figure*}
\begin{tabular}{ccc}
\multirow{2}{*}[0in]{\includegraphics[width=0.3\hsize]{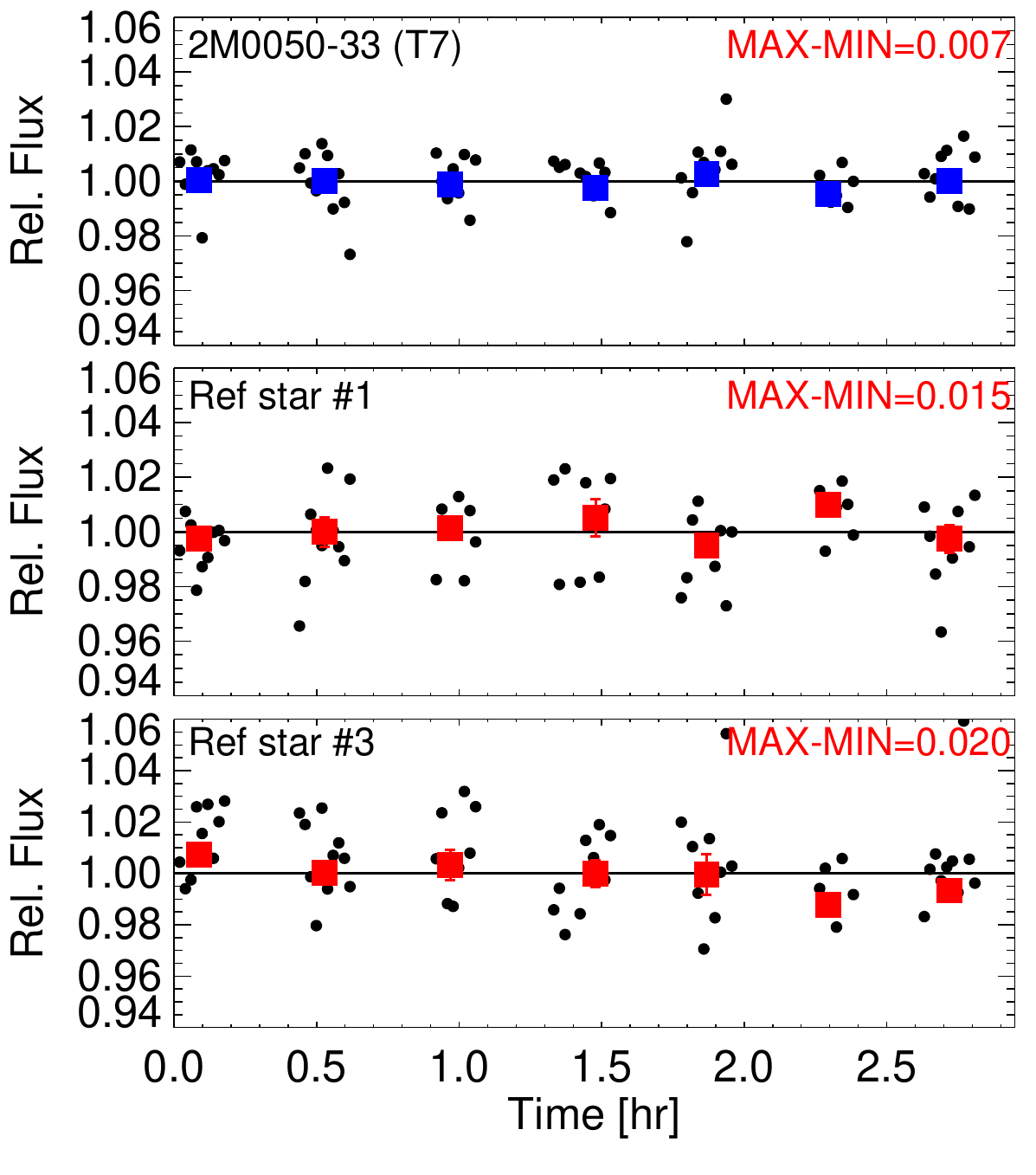}} &
\multirow{2}{*}[0in]{\hspace{-0.2 in} \includegraphics[width=0.3\hsize]{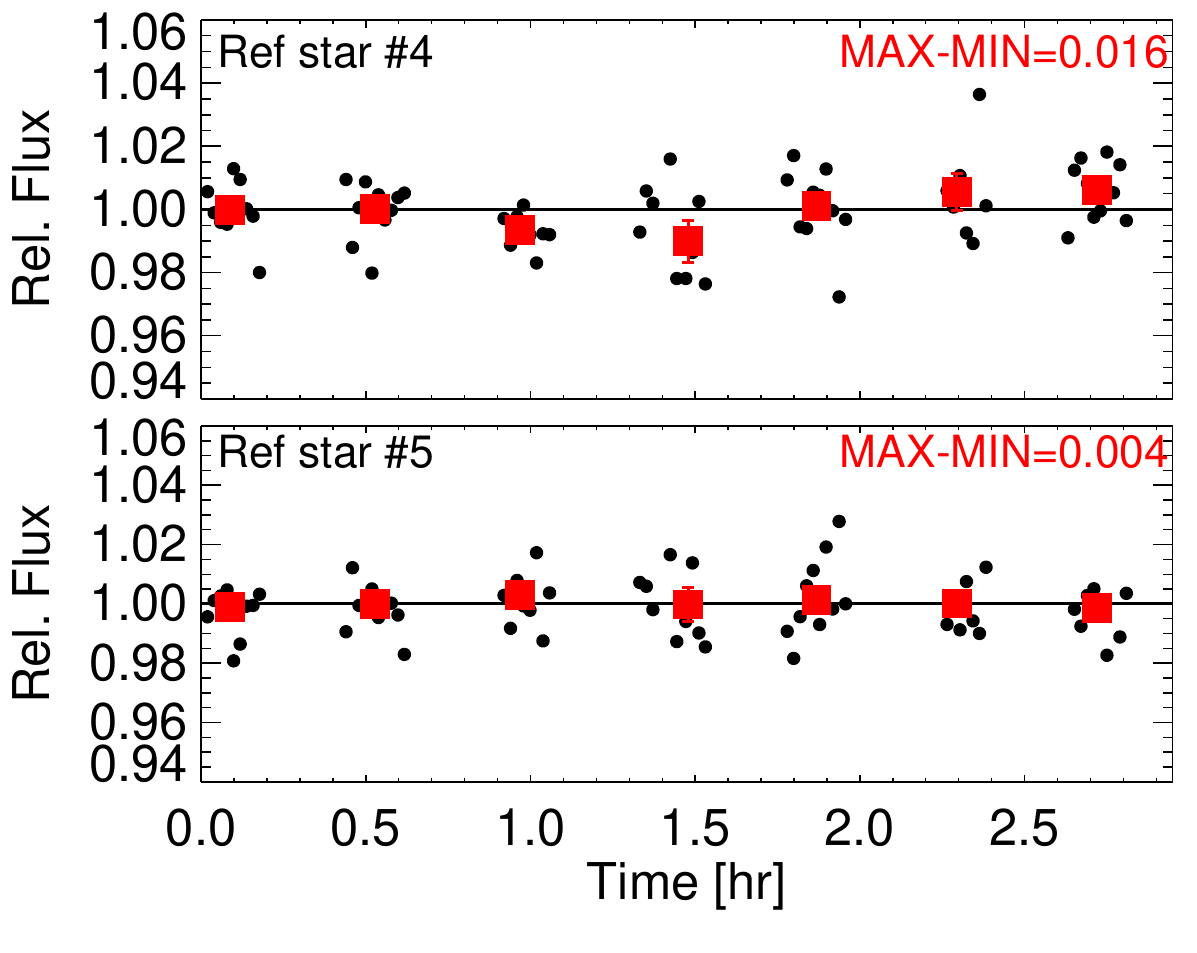}} &
\multirow{1}{*}[0.15in]{\hspace{-0.25in}\includegraphics[width=0.36\hsize]{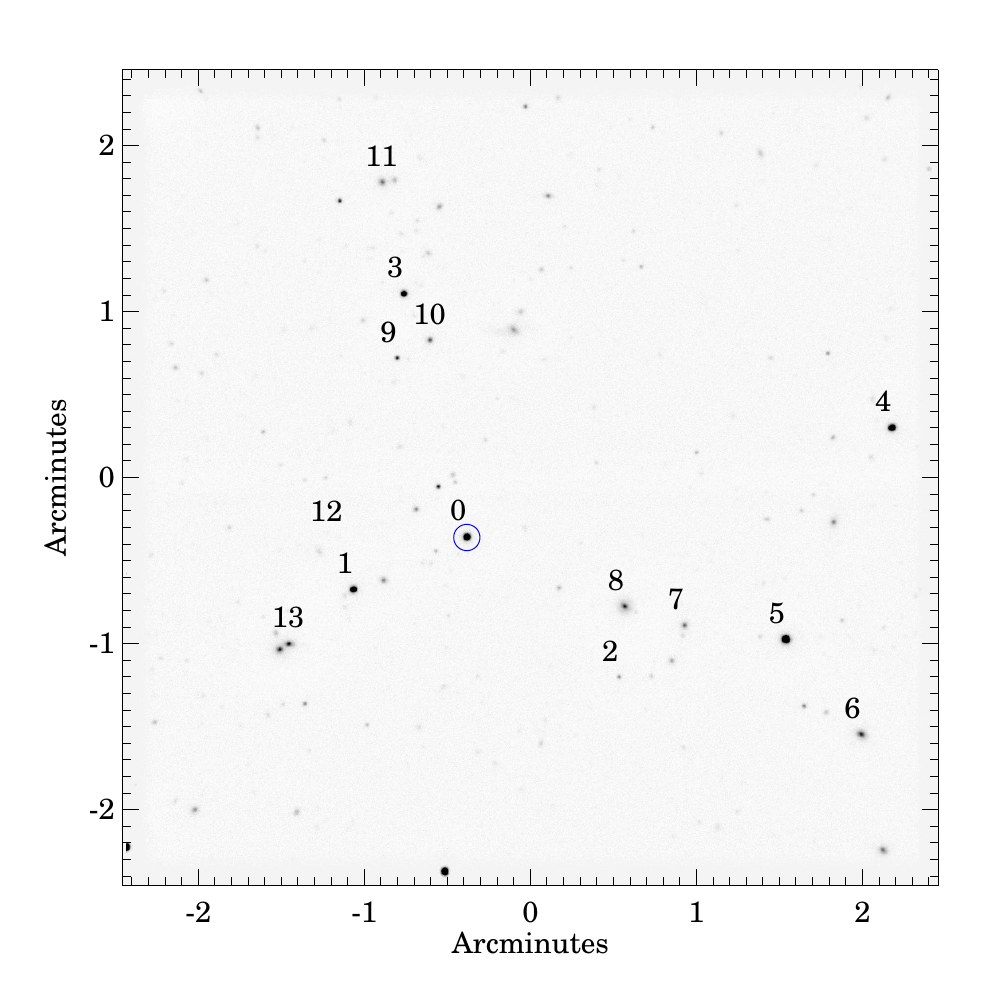}} 
\end{tabular}
\vspace{2.3 in}\caption{Light curve for the T7 dwarf 2M0050-33, and four reference stars.  Binned data points are represented as filled squares (blue for target, and red for references), while high cadence photometry from individual exposures are plotted as black points.  The reference stars are labelled by their number in the finding charts in the rightmost panel.  The target is always labelled with index `0' and is circled.  The MAX-MIN values of the binned light curves are shown in the top right corner of each panel. W14 report a target amplitude of 10.8$\pm$1.3\% for this time series.\label{fig:lc1}}
\end{figure*}

 \begin{figure*}
\begin{tabular}{ccc}
\multirow{2}{*}[0in]{\includegraphics[width=0.3\hsize]{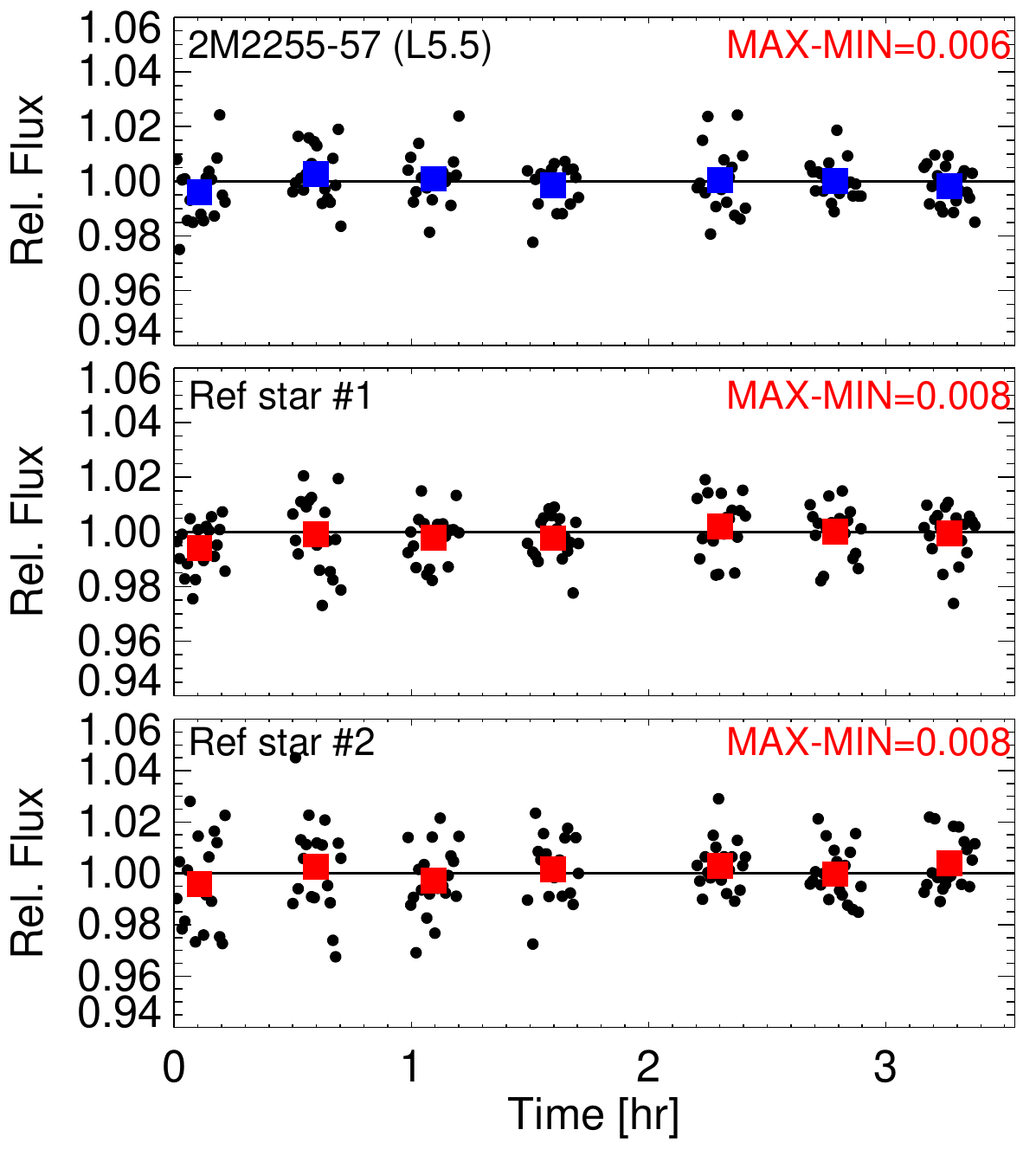}} &
\multirow{2}{*}[0in]{\hspace{-0.2 in} \includegraphics[width=0.3\hsize]{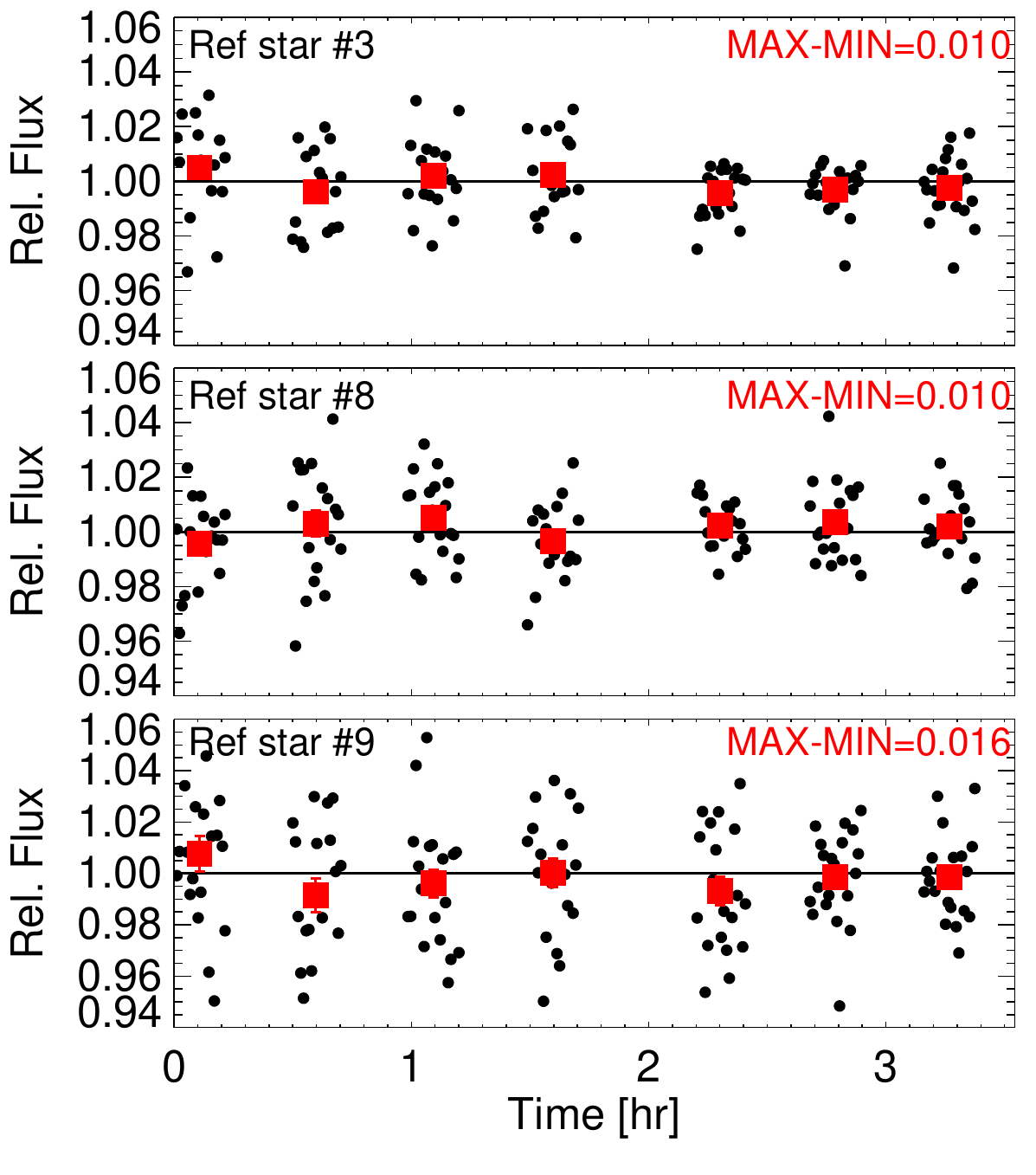}} &
\multirow{1}{*}[0.15in]{\hspace{-0.25in}\includegraphics[width=0.36\hsize]{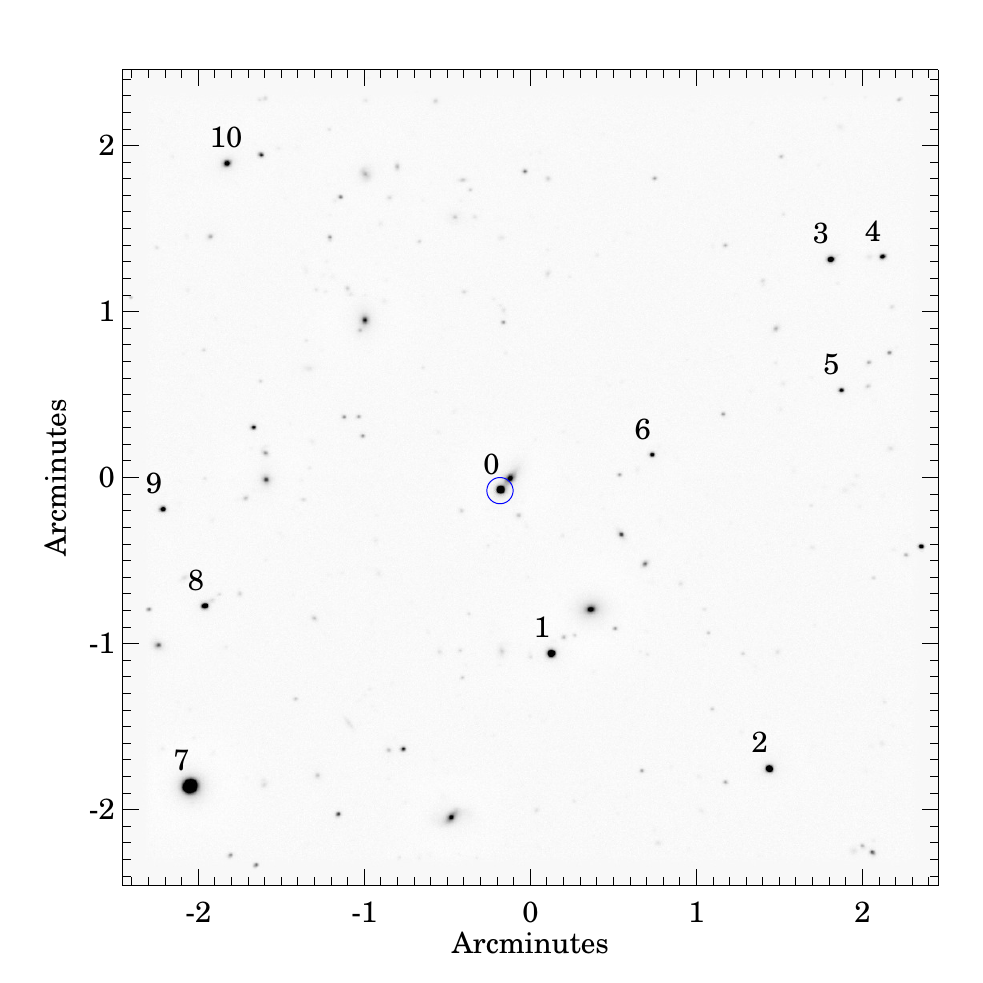}} 
\end{tabular}
\vspace{2.3 in}\caption{Same as figure \ref{fig:lc1} but for the L5 dwarf 2M2255-57, and five reference stars.  W14 report a target amplitude of 9.4$\pm$1.6\% for this time series. \label{fig:lc2}}
\end{figure*}

\begin{figure*}
\begin{tabular}{ccc}
\multirow{2}{*}[0in]{\includegraphics[width=0.3\hsize]{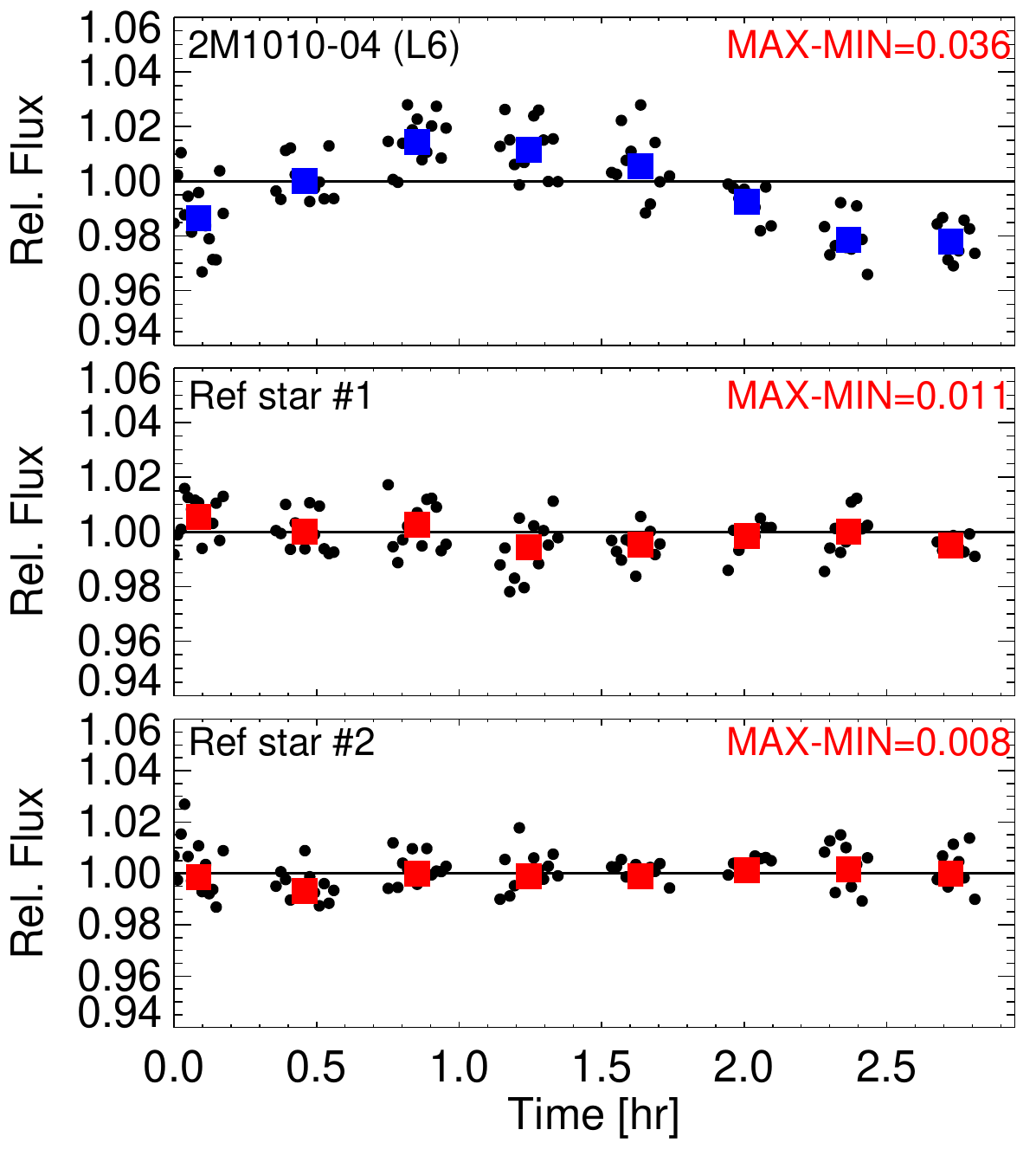}} &
\multirow{2}{*}[0in]{\hspace{-0.2 in} \includegraphics[width=0.3\hsize]{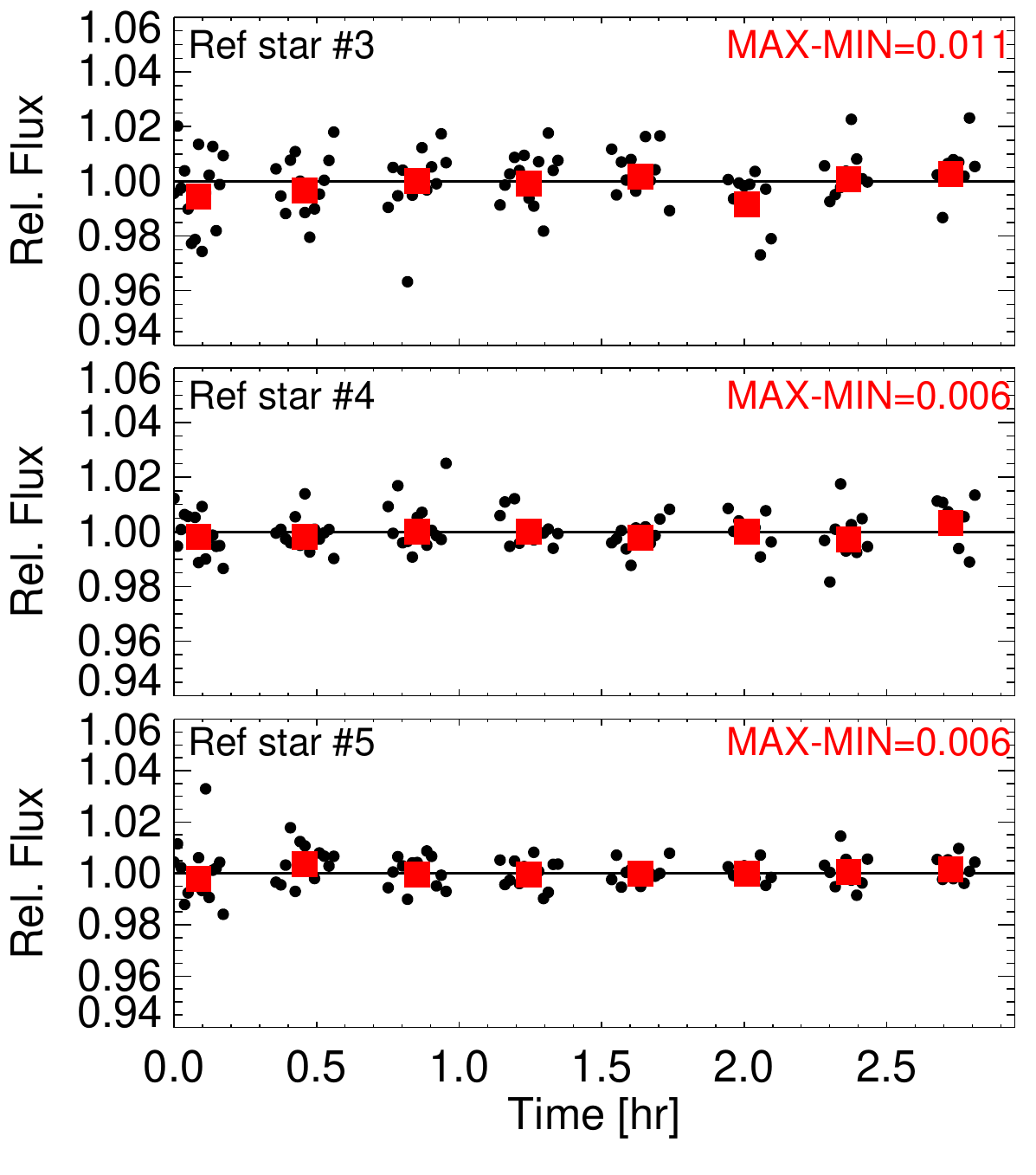}} &
\multirow{1}{*}[0.15in]{\hspace{-0.26in}\includegraphics[width=0.36\hsize]{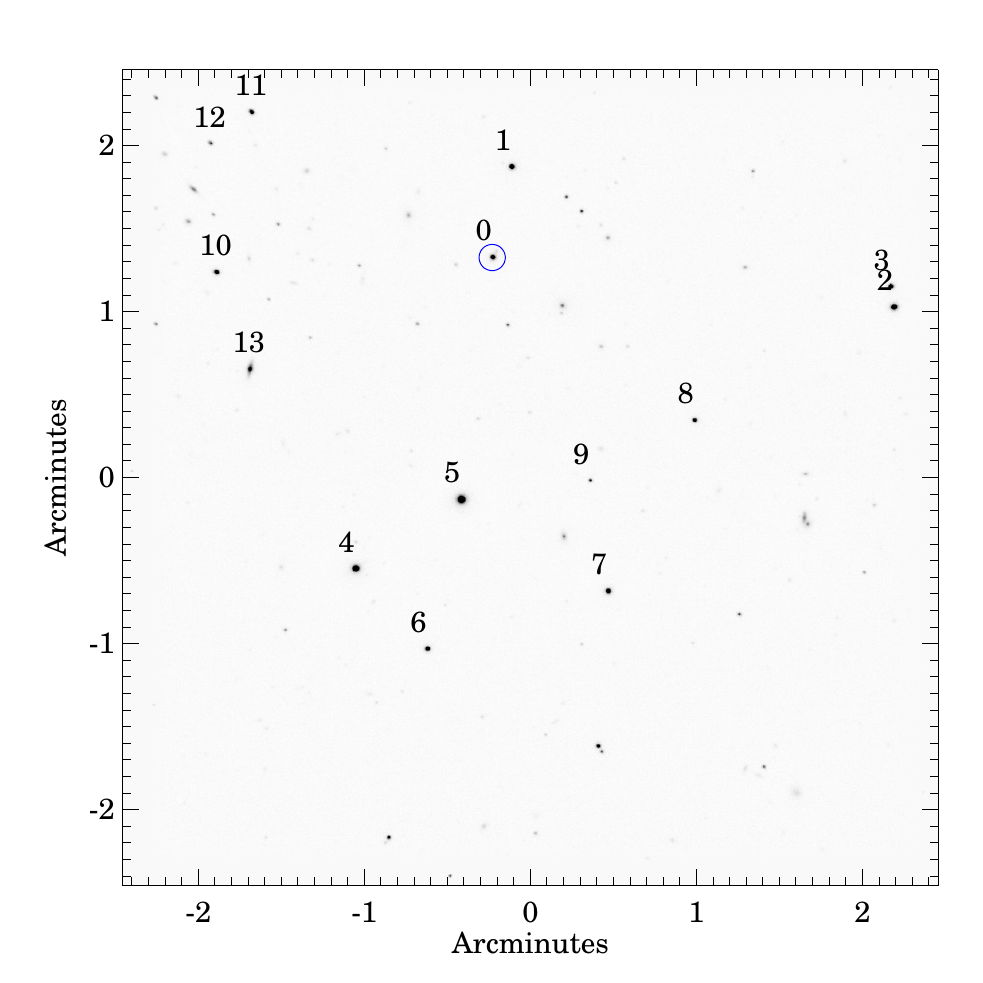}} 
\end{tabular}
\vspace{2.3 in}\caption{Same as figure \ref{fig:lc1} but for the L6 target 2M1010-06, and five reference stars.  W14 report a target amplitude of 4.3$\pm$1.2\% for this time series.\label{fig:lc3}}
\end{figure*}

\section{Results}
\label{sect:results}
We reduced archived time series data for 14 targets reported to be variable with high significance (p$<$0.05 according a $\chi^2$ test) by W14, 13 of which are reported to vary with amplitudes of $\sim$2\%-11\%.  Relative flux light curves were generated for each target and up to 5 well-behaved reference stars used to calibrate the target light curve.  As opposed to stacking images in each $\sim$15 minute epoch as done in the BAM analysis, we extracted photometry from individual images, and subsequently binned points together to make an appropriate comparison.  Example light curves showing both high-cadence and binned data for 3 targets (two non-detections, and one detection) are shown in figures \ref{fig:lc1}-\ref{fig:lc3}, while the remaining light curves are provided in the Appendix.  Following the example of W14 we have measured peak-to-trough amplitudes from the binned light curves by taking the difference between their maximum and minimum brightness.  Since correctly identifying the target is crucial to any time series analysis, finding charts labeling the target and reference star positions are also provided.  

For 10 of 14 targets we find no evidence for variability above the  $\sim$1.0\%-1.6\% level.  The remaining four targets display variations $>$2\% at a level exceeding those of similar-brightness reference stars.    Of these, two are previously known large-amplitude variables in the $J$ band:  the early T-dwarfs SIMP 0136 and 2M2139 \citep{artigau09,radigan12,khandrika13,apai13}.  Another, the T6.5 dwarf 2M2228, is a known large-amplitude variable at redder wavelengths \citep{buenzli12} and has been previously observed to vary in the $J$ band at a level of $\sim$1.4\%-1.8\% \citep{clarke08,buenzli12,radigan14}.  Finally, we confirm W14's identification of the L6 dwarf 2M1010 as a new large-amplitude variable, for which we  measure a peak-to-trough amplitude of 3.6\%.

Two targets, 2M1126 and 2M0835, show evidence of lower level variations ($\sim$1.6\% and  $\sim$1.3\%) exceeding those of similar-brightness references, the latter agreeing (within uncertainties) with W14's reported amplitude of 1.7$\pm$0.5\%.  Weak variability of 2M1126 is also reported in R14, with an amplitude of 1.2\%.  Amplitudes and upper limits determined for all targets in our analysis are provided in table \ref{tab:targets}.  In cases where the target and reference stars vary by similar amounts, amplitudes are reported as upper limits.  For targets displaying trends at a level exceeding those of similar-brightness references by a factor of $\sim$2, a tentative variability amplitude is assigned, but since our goal in this paper is to place limits on large-amplitude variability, we make no further attempt to determine the significance of lower-level trends. 


\begin{table*}
\centering{
\footnotesize
\begin{threeparttable}
\caption{Variable L and T dwarfs reported by W14 ($p<0.05$) \label{tab:targets}}
\begin{tabular}{lcccccc}
\hline
\hline
Target Name    &   Opt   &	NIR   &   $J_{\rm 2M}$ 	 &    Refs &     \multicolumn{2}{c}{SofI Amplitudes}  \\
			&    SpT  & 	SpT   &       (mag)             &             &   W14 & This paper        \\
			&            &                 &                              &             &        (\%)     &  (\%)                         \\	

 \multicolumn{7}{c}{New large-amplitude variables reported by W14} \\
 \hline
2MASS J00501994-3322402 	& \nodata & T7  	&  15.93 $\pm$ 0.07   &  T05,B06 & $10.8\pm1.3$  & $<0.7\pm0.4$\\
2MASS J01062285-5933185   &  L0 	& \nodata 	&   14.33 $\pm$ 0.04   &  R08         &$4.3\pm1.2$     & $<1.2\pm0.4 $\\
2MASS J03480772-6022270 	& \nodata & T7 	&   15.32 $\pm$ 0.05   &  B03,B06         & $2.4\pm0.5$    & $<1.1\pm0.4$\\
2MASS J03582255-4116060 	&  L5        & \nodata &    15.85 $\pm$ 0.09  &  R08       & $4.8\pm1.2$    & $<0.7\pm0.8$\\
2MASS J04390101-2353083 	&  L6.5     & \nodata &    14.41 $\pm$ 0.03  &   C03        & $2.6\pm0.5$    & $<1.2\pm0.3$\\		  
2MASS J10101480-0406499 	&  L6        & \nodata	&    15.51 $\pm$ 0.06  &  C03         &  $5.1\pm1.1$   & $3.6\pm0.4$\\
2MASS J11263991-5003550 	&  L4        & L6.5 	&    14.00 $\pm$ 0.03  &  F07,B08  & $3.2\pm0.7$    & $1.6\pm0.2$ \\
2MASS J12074717+0244249	&  \nodata  & T0 	& 15.58 $\pm$ 0.07     &  H02,B08         &  $5.2\pm1.1$   & $<1.5\pm0.4$\\
2MASS J13004255+1912354 	&  L1        & L3 	        &     12.72 $\pm$ 0.02 &   G00,B08 & $9.6\pm0.9$   & $<0.7\pm0.3$\\
2MASS J22551861-5713056\tablenotemark{b} 	& \nodata & L5.5 	&     14.08 $\pm$ 0.03 &  K07        & $9.4\pm1.6$   & $<0.6\pm0.3$\\

\multicolumn{7}{c}{New weak variables reported by W14}\\
\hline
2MASS J08354256-0819237 	&  L5        & \nodata 	&    13.17 $\pm$ 0.02  &   C03        & $1.7\pm0.5$    & $1.3\pm0.2$\\   
\multicolumn{7}{c}{Previously known variables reported by W14}\\
\hline
 SIMP J013656.5+093347.3 	& \nodata & T2.5 	&   13.46 $\pm$ 0.03   &  A06,A09         &$3.0\pm0.6$     &  $3.2\pm0.4$\\
 2MASS J21392676+0220226 	& T0         & T1.5     &      15.26 $\pm$ 0.05 &  R08,R12        & $4.7\pm0.5$   & $5.9\pm0.4$\\
 2MASS J22282889-4310262	& \nodata & T6.5 	&     15.66 $\pm$ 0.07 &  B03,B12         & $3.9\pm0.7$   & $2.9\pm0.7$\tablenotemark{a} \\ 
 \hline
\end{tabular}
\begin{tablenotes}
\vspace{1mm}
\item \tablerefs{\citet{burgasser03_opt}[B03], \citet{burgasser06}[B06], \citet{burgasser08}[B08], \citet{cruz03}[C03], \citet{gizis00} [G00], \citet{folkes07}[F07], \citet{reid08}, \citet{hawley02}[H02], \citet{kendall07}[K07], \citet{tinney05}[T05]}
\vspace{1mm}
\item $^{\rm a}$This target is a known weak variable in the $J$ band with an amplitude of $\sim$1.4\%-1.8\% \citep{clarke08,radigan14,buenzli12}, and is known to vary with larger amplitude at longer wavelengths \citep{buenzli12}.   
\item $^{\rm b}$This target is a resolved binary, and excluded from the statistical analysis presented in figure \ref{fig:key} and table \ref{tab:stats}.
\end{tablenotes}
\end{threeparttable}}
\end{table*}

\begin{table*}
\centering{
\footnotesize
\begin{threeparttable}
\caption{Observed frequency of large-amplitude $J$-band variability as a function of spectral type \label{tab:stats}}
\label{tab:var_freq}
\begin{tabular}{cccccccccc}
\hline
\hline
Spectral Type   &   \multicolumn{3}{c}{rBAM~~~~~~~~~}   &	\multicolumn{3}{c}{R14~~~~~~~~~}   &   \multicolumn{3}{c}{rBAM$+$R14~~} \\
 Range &  N  & n & ~~~freq. (\%)~~~~~     &  N  & n & ~~~freq. (\%)~~~~~      &  N  & n & ~~~freq. (\%)~~~~~  \\
\hline	
\vspace{-1mm}& & & & & & & & \\	 
\vspace{2mm}
L0-L8.5             & 27  &   1 & 3.7$^{+5.0}_{-2.6}$   	  & 15  &   0 & 0.0$^{+6.8}_{-0.0}$   	  & 34  &   1 & 2.9$^{+4.1}_{-2.1}$  \\ 
\vspace{2mm}
L9-T3.5		  & 8  &   2 & 25$^{+15}_{-12}$   	  & 15\tablenotemark{a}  &   4 & 27$^{+12}_{-10}$   	  & 17  &   4 & 24$^{+11}_{-9}$   \\

\vspace{2mm}
T4-T9.5		  & 23  &   1 & 4.3$^{+5.8}_{-3.1}$   	  &   26 &   0 & 0.0$^{+4.1}_{-0.0}$     & 31  &   1 & 3.2$^{+4.4}_{-2.3}$   \\
\hline		
\vspace{-2mm}& & & & & & & & \\	
\vspace{2mm}		 
L0-L8.5 $+$ T4-T9.5 & 50 & 2 & 4.0$^{+3.5}_{-2.2}$ & 41 &  0 & 0.0$^{+2.7}_{-0.0}$ & 65 & 2 & 3.1$^{+2.7}_{-1.7}$ \\
\vspace{1mm}
All & 58 & 4 &  6.9$_{-2.8}^{+3.8}$ & 56 & 4 &  7.1$_{-2.9}^{+3.9}$ & 82 & 6 &  7.3$_{-2.5}^{+3.2}$\\
\hline		
\end{tabular}
\begin{tablenotes}
\item \tablecomments{N=number of objects observed, n=number variable above 2\%. Uncertainties correspond to shortest 68\% Binomial credible intervals. rBAM refers to the revised BAM sample after taking into account non-detections reported in this work, and excluding resolved binaries from the statistical sample as done in R14. }
$^a$This number has been revised downward from 16 in R14, due to the T0 dwarf SDSS J151114.66+060742.9 being resolved into a binary (Gelino et al., in prep) with estimated component types of L5.5 and T5 \citep{burgasser10}.
\end{tablenotes}
\end{threeparttable}}
\end{table*}

\section{Discussion: the frequency of large-amplitude variability}
\label{sect:discuss}

\begin{figure}[ht!]
 \includegraphics[width=0.9\hsize]{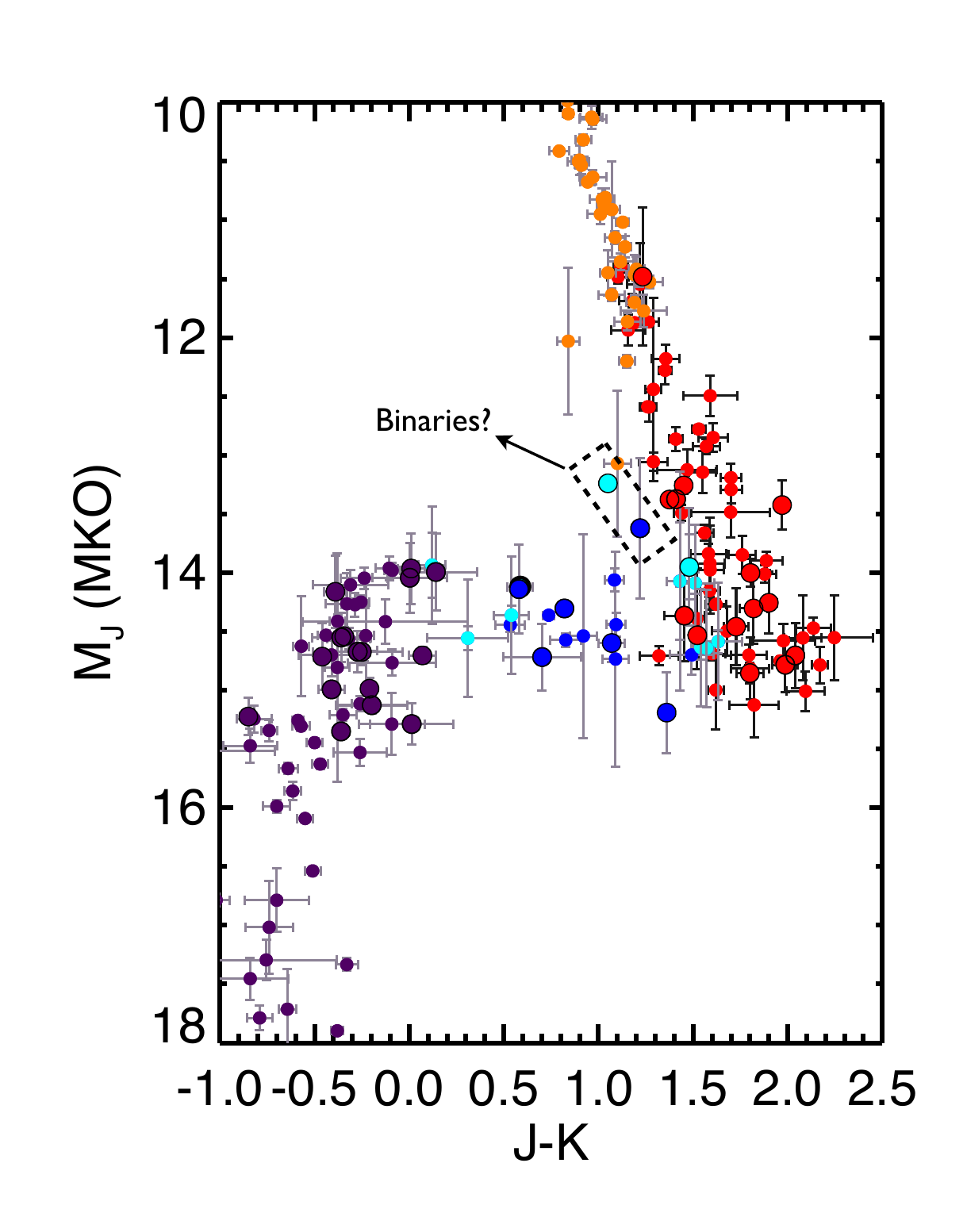} 
\caption{ Sample composition of rBAM and R14 survey targets with parallaxes (large circles) shown alongside all M, L, and T dwarfs with parallaxes (small circles) from the database of T. J.  Dupuy.  The points are divided by color into the spectral type bins of R14: L-dwarfs ($\le$ L8.5) in red, T-dwarfs ($\ge$ T4) in purple,  and L/T transition dwarfs (L9-T3.5) in dark and light blue.  Within the L/T transition bin T0-T2.5 dwarfs are shown in dark blue, while objects with L9-L9.5 and T3-T3.5 spectral types (the endpoints) are plotted in a lighter color to illustrate the degree of overlap with the L-dwarf and T-dwarf bins. Two L/T transition objects in the rBAM+R14 sample are overluminous for their spectral types (L9 and T0) are noted as possible binaries. \label{fig:cmd}}
\end{figure}

\begin{figure*}[ht!]
 \includegraphics[width=1\hsize]{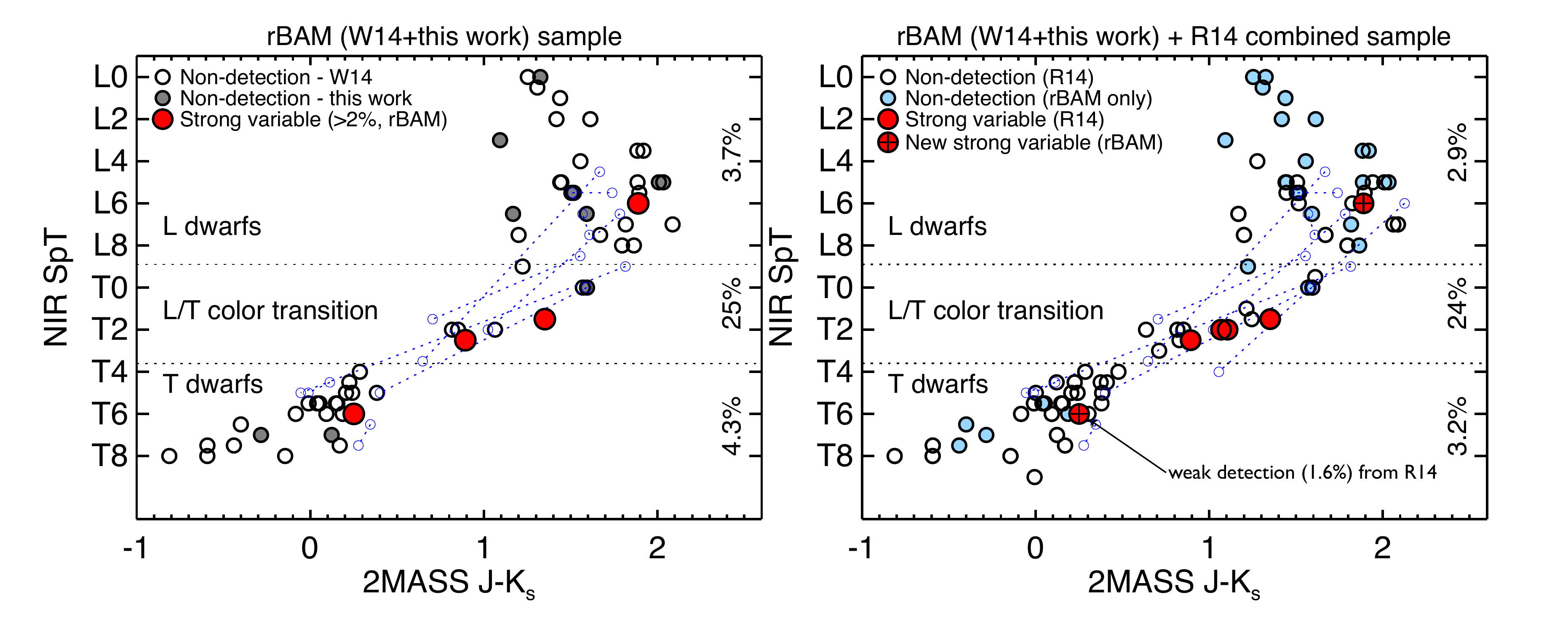} 
\caption{Sample composition of rBAM (left) and rBAM$+$R14 (right, with the difference sample highlighted in blue) variability surveys.   Large-amplitude variables have been identified with filled red circles.  The percentages on the right axes give the observed fraction of large-amplitude variables in each spectral type bin.   Binary components from both W14 and R14 (not included in our analysis; see section \ref{sect:sample}) are shown as blue open circles connected by dotted lines.  \label{fig:key}}
\end{figure*}

Our independent analysis of NTT/SofI data from the BAM program finds evidence for variability $>$2\% in 4 of 13 targets originally reported as large-amplitude variables.  
Here we examine how the re-classification of variables and amplitudes in the BAM sample affects the inferred frequency of large-amplitude variability overall, and as a function of spectral type.   In the following discussion non-detections reported by W14 are assumed to be reliable.  Although we have not independently analyzed the reported non-detections, we reason that obtaining a false negative for large-amplitude variability intrinsic to a given target is unlikely.  

\subsection{rBAM: a Revised BAM Sample}
\label{sect:sample}
Combining our result with additional non-detections reported by W14, we can define a revised BAM sample (rBAM, hereafter) from which occurrence rates of large-amplitude variability can be examined.  There are 55 non-detections of variability reported by W14, yielding a total of 65 non-variable targets, and 4 variables at amplitude levels $>$2\%.  Of the 69  BAM targets, 13 have been identified as resolved binaries in high angular resolution images (see tables 1 and 2 of W14\footnote{In addition to the resolved systems reported in tables 1 and 2 of W14, the T0 dwarf SDSS J151114.66$+$060742.9---which is a strong binary candidate of \citet{burgasser10} with L5.5 and T5 components, and over-luminous in the parallax study of \citet{faherty12}---has been resolved into binary (Gelino et al., in prep).}).  The binary components are distributed in spectral type as follows: 12 L4-L8 dwarfs, 5 L9-T3 dwarfs, and 9 T4-T8 dwarfs.   The inclusion of known binaries in an unbiased sample poses several challenges:

\begin{itemize}

\item The unresolved spectral type of a binary system may not be representative of the individual components.  Thus individual components must be considered as separate targets.  This is particularly true of L/T transition binaries for which frequently only one or neither of the components are transition objects themselves.  

\item If an unresolved system is found to be variable, there is no way of knowing which component is responsible.  If the components have significantly different spectral types, to which spectral type should the detection be assigned?

\item  Variability is more difficult to detect in binary systems.  The amplitude required for an individual component to produce a system amplitude of $A$ is $A^{\prime} = (1+10^{0.4\Delta m})A$, where $\Delta m$ is the contrast ratio of the binary.  Sensitivity to large-amplitude variations ($>$2\%) consequently requires sensitivity to system amplitudes of $\sim$1\%  (and $<$1\% in the case of L9-T3.5 components in the BAM sample, which have fluxes less than or equal to those of their companions).  Since percent-level signals are not routinely detectable, detections of large-amplitude variability for binary systems will be highly incomplete.

\end{itemize}

Given these complications, binaries were excluded from the statistical analysis of R14.  An exception was made for the (non-variable) high flux ratio binary 2M1209, wherein the T2.5 primary dominates the total flux ($\Delta m\sim1.5$).  For this target most large-amplitude signals from the T2.5 dwarf would have been detectable ($A^{\prime}/A\sim1.25$ ), whereas signals from the secondary would have been largely washed out ($A^{\prime}/A\sim5$ ).  We have followed a similar approach with the rBAM sample, excluding known binaries except in two cases mirroring that of 2M1209, where component flux ratios were large: the L6 binary 2M2255 ($\Delta m\sim1.5$)  and the T5.5/T8 binary 2M1225 ($\Delta m\sim1.3$). The effect of including binaries in the sample is considered in section \ref{sect:combined}.  

For the purpose of examining variability as a function of spectral type, we have adopted the spectral type bins of R14, which divides the sample into 3 groups: L-dwarfs\footnote{Note that the R14 sample did not extend to types earlier than L4.} (L0-L8.5), the L/T transition (L9-T3.5) and T-dwarfs (T4-T9).  The boundaries of the L/T transition bin select objects within the NIR color reversal connecting the L-dwarf and T-dwarf branches of the NIR CMD \citep[c.f. figure 16 of][]{dupuy12}; a region which was largely missed by early searches for NIR variability.  A CMD of the rBAM sample and adopted spectral type bins is shown in figure \ref{fig:cmd}.  We note that two objects included in the L/T transition sample appear over-luminous on the CMD, and may be binary contaminants (however, we have not excluded any objects on the basis of {\it suspected} binarity).  Excluding resolved binaries, the sample contains 58 unique targets: 27 L0-L8.5, 8 L9-T3.5 dwarfs, and 23 T4-T9 dwarfs.

\subsection{Large-Amplitude Variability in the rBAM Sample}
Here we examine observed occurrence rates of variability above 2\% in the rBAM sample.  The 2\% threshold is adopted from R14, and informed by previous monitoring campaigns that have found brown dwarf variability above this level to be rare \citep{koen04a,koen05a,clarke08,girardin13} (e.g. see section \ref{sect:intro}).  In addition, according to simulations by W14, it is the amplitude at which variability is broadly detectable for the BAM targets.

In the rBAM sample, the overall fraction of targets found here to vary with peak-to-trough amplitudes $>2$\% is 4/58 or 6.8$_{-2.8}^{+3.8}$\%.  Uncertainties have been estimated from 68\% Binomial shortest confidence intervals.\footnote{We caution that the assumption of Binomial statistics does not hold when the signal sensitivity varies target to target.  However, because we are restricting our analysis to large-amplitude signals that would have been detectable in most target light curves, the Binomial confidence interval provides a reasonable (even if not strictly accurate) estimate of the counting uncertainties.}  This number is revised downward from a frequency of 13/58  or $22^{+6}_{-5}$\% based on large-amplitude detections reported in W14.  As a function of spectral type, large-amplitude variability is observed for 1/27  (3.7$^{+5.0}_{-2.6}$\%) L0-L8 dwarfs, 2/8 (25$^{+15}_{-12}$\%) L9-T3 dwarfs and 1/23 (4.4$^{+5.8}_{-3.1}$\%) T4-T9 dwarfs, consistent with the finding of R14 that large-amplitude variability occurs more frequently at the L/T transition.  If we consider only objects outside the L/T transition 2/50 or $3.9^{+3.4}_{-2.1}$\% are large-amplitude variables, suggesting that strong variability of ultracool dwarfs is rare outside of L9-T3.5 spectral types.

\subsection{Variability in a combined rBAM$+$R14 sample}
\label{sect:combined}
A larger sample can be considered by combining the rBAM and R14 results together.  The rBAM sample is complementary to that of R14, filling in L0-L3 spectral types, and providing a larger sample of L4-L8 dwarfs with varied $J-K_s$ colors.  The combined sample yields 83 unique targets with L0-T9 spectral types (also shown in figure \ref{fig:key} and table \ref{tab:stats}).   Reported photometric precision and observation lengths are comparable for both surveys.  Known binaries have been excluded from both samples (with 3 high-flux ratio systems excepted, as described in section \ref{sect:combined}). The overall frequency of large-amplitude variability in the combined sample is 6/82 or $7.3^{+3.2}_{-2.5}$\%.

Outside the L/T transition, 2/65 (3.1$_{-1.7}^{+2.7}$\%) targets are observed to be high-amplitude variables, compared to 4/17 (24$^{+11}_{-9}$\%) of targets with L9-T3.5 spectral types.  To determine the significance of this difference we performed a simulation of 100,000 trials for which 6 objects (corresponding to the six large-amplitude variables) are drawn randomly from a sample containing 17 objects of type 1 (mirroring the L/T transition sample) and 65 objects of type 2 (mirroring the non-L/T transition sample).   Four or more objects of type 1, as we have observed, are drawn by random chance in only 1.5\% of trials.\footnote{A two-proportion z-test of the two population yields $p<0.007$, but overestimates the significance due to the small sample sizes involved.  An exact test for two binomial distributions is described in the Appexdix of \citet{brandeker06}, and yields a result identical to that of our simulation.}  Thus, observational evidence to date favors a an increase in the frequency of large-amplitude variability for L9-T3.5 dwarfs in comparison to other spectral types, at the  98.5\% confidence level.   Including binary components in the sample (counted as either half or whole objects) does not change the significance of our findings.  In addition, one of the large-amplitude rBAM variables outside the L/T transition (the T6.5 dwarf 2M2228) was observed to vary with a lower amplitude of 1.6\% in the survey of R14\footnote{This source was also found to vary with an amplitude of 1.4\% by \citet{clarke08}.}, and if occurrences of large amplitude-signals are counted on a per-visit rather than a per-target basis, the significance of our conclusions are strengthened (p=0.0007 or $>$99.7\% confidence).

Multi-wavelength monitoring of variable L/T transition brown dwarfs \citep{artigau09,radigan12,apai13,burgasser14} has provided constraints on the types of surface patches involved. Comparisons with model atmospheres \citep{allard03,marley02,tsuji03,burrows06} have shown that cool regions of high condensate opacity interspersed with warm regions of lower condensate opacity are required to reproduce the observations.  This is consistent with the expectation that regions of lower condensate opacity will naturally appear brighter/warmer in the $J$-band, due to the $\tau=2/3$ surface sitting deeper in the photosphere. The variability amplitude resulting from an atmosphere composed of two surface components is a function 3 parameters: the mean filling factor of the warmer surface component, $\left< a \right>$, the change in filling factor, $\Delta a$, and the contrast in flux (or equivalently brightness temperature) between components $f_{\rm hot}/f_{\rm cold}$ \citep[e.g. see expressions in][]{radigan12,burgasser14},

\begin{equation}
\frac{\Delta F}{\left < F \right >} = \frac{\Delta a}{\left< a \right>+\epsilon}
\end{equation}

\noindent where $\left<F\right>=\left< a \right> f_{\rm hot} + (1-\left< a \right>)f_{\rm cold}$ and $\epsilon=1/(f_{\rm hot}/f_{\rm cold}-1)$. 
Therefore, large-amplitude variability at the L/T transition may be caused by changes in the spatial scales of cloud features ($\propto \Delta a$), and/or increasing contrast ($\propto f_{\rm hot}/f_{\rm cold}$) between features as clouds dissipate.  All else being equal, as the average coverage of warm regions increases ($\propto \left< a \right>$), observed amplitudes will decrease, favoring larger amplitudes for objects at the beginning and middle stages of cloud dissipation over those at the end stages.

\section{Summary and Conclusions}

We presented an independent reduction and analysis of SofI/NTT time series data for the 14 L and T dwarfs reported to be significantly variable by W14, 13 of which are reported to vary with amplitudes $>$2\%.   We find evidence of large-amplitude variability for 4 of these targets, and place upper limits of $\sim$1\%-\%-1.6\% on variability of the remaining sample (see table 1). For two targets we find evidence of weak variability at amplitudes of 1.3\% and 1.6\%.  Of the 4 targets for which we confirm strong variability, 3 are well-documented variables in the literature (the early T-dwarfs SIMP0136 and 2M2139, and T6.5 dwarf 2M2228), while one is newly identified in W14 (the L6 dwarf 2M1010), although our analysis finds a slightly lower amplitude.

Based on our revised classification of variable targets in the BAM sample we find large-amplitude variability of brown dwarfs to be rare at NIR wavelengths, with the exception of the L/T transition.  In a combined sample of 82 unique L and T dwarfs from the revised BAM sample and R14,  we find an overall observed frequency for large-amplitude variability of $7.3^{+3.2}_{-2.5}$\%.    Outside the L/T transition the frequency of large-amplitude variability is 3.2$_{-1.8}^{+2.8}$\%, in comparison to 24$^{+11}_{-9}$\% for L9-T3.5 spectral types.  The null hypothesis of equal occurrence rates inside and outside the L/T transition is ruled out at a confidence level of  98.5\%.  Therefore, while variability with amplitudes $>$2\% is not exclusive to the L/T transition, there is compelling observational evidence to suggest it is more common at these spectral types, indicative of larger spatial scales and/or $J$-band contrasts between surface features at these spectral types.  

While the surveys discussed above provide snapshots of variability at single epochs, amplitudes and light curve morphologies for variable brown dwarfs are known to evolve with time \citep{artigau09,radigan12,metchev13_proc,gillon13}.  Therefore, long-term follow up of  variables  both inside and outside of the L/T transition should provide valuable insight into the time-averaged properties of variability as a function of spectral type, and shed light on the dynamic timescales for variability across the L-T sequence \citep{xi14}.

\label{sect:conclusions}

\clearpage

\appendix
\section{A. Light curves and finding charts for remaining targets and reference stars}

\vspace{-3mm}\begin{figure*}[ht!]
\begin{tabular}{ccc}
\multirow{2}{*}[0in]{\includegraphics[width=0.3\hsize]{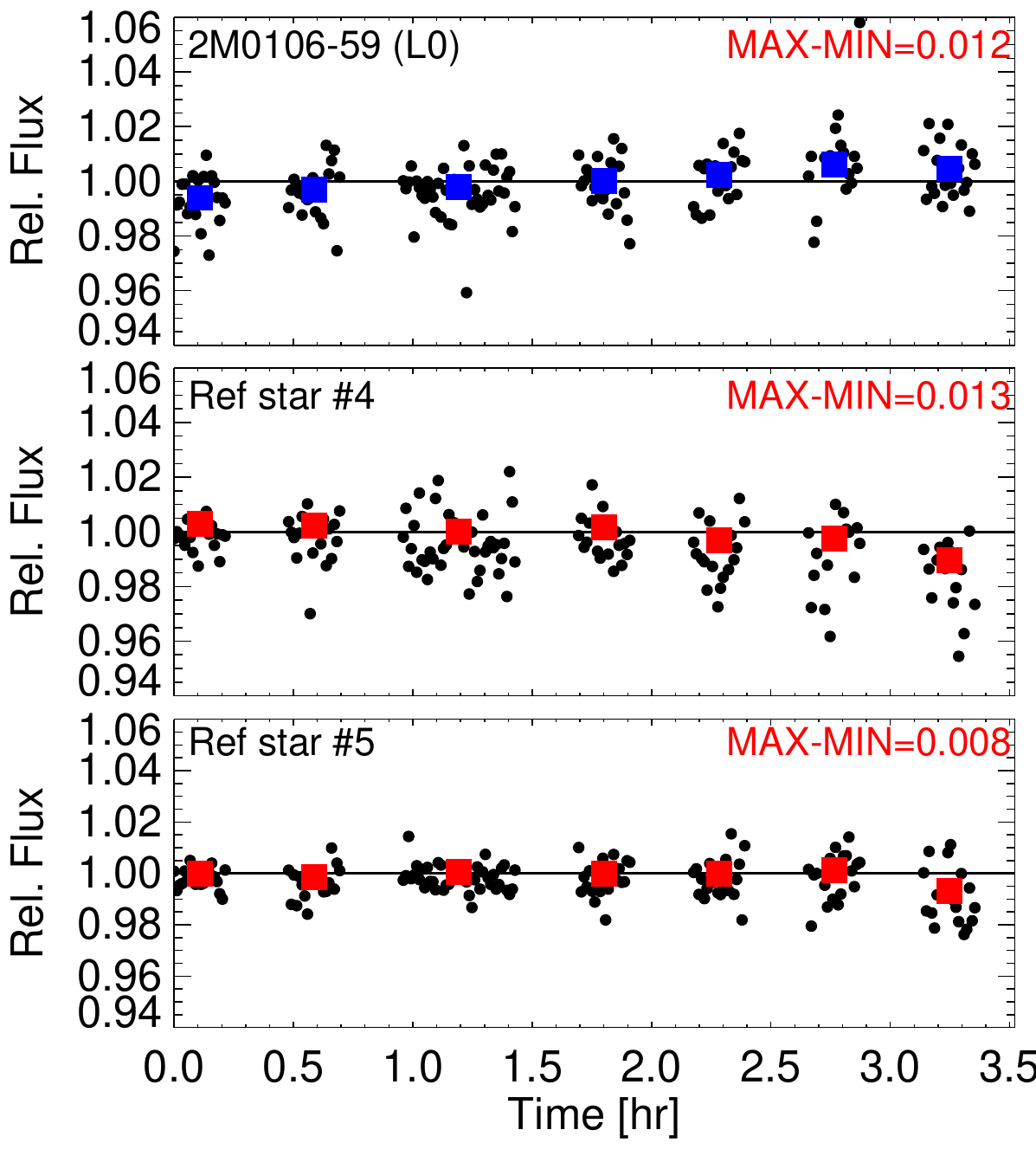}} &
\multirow{2}{*}[0in]{\hspace{-0.2 in} \includegraphics[width=0.3\hsize]{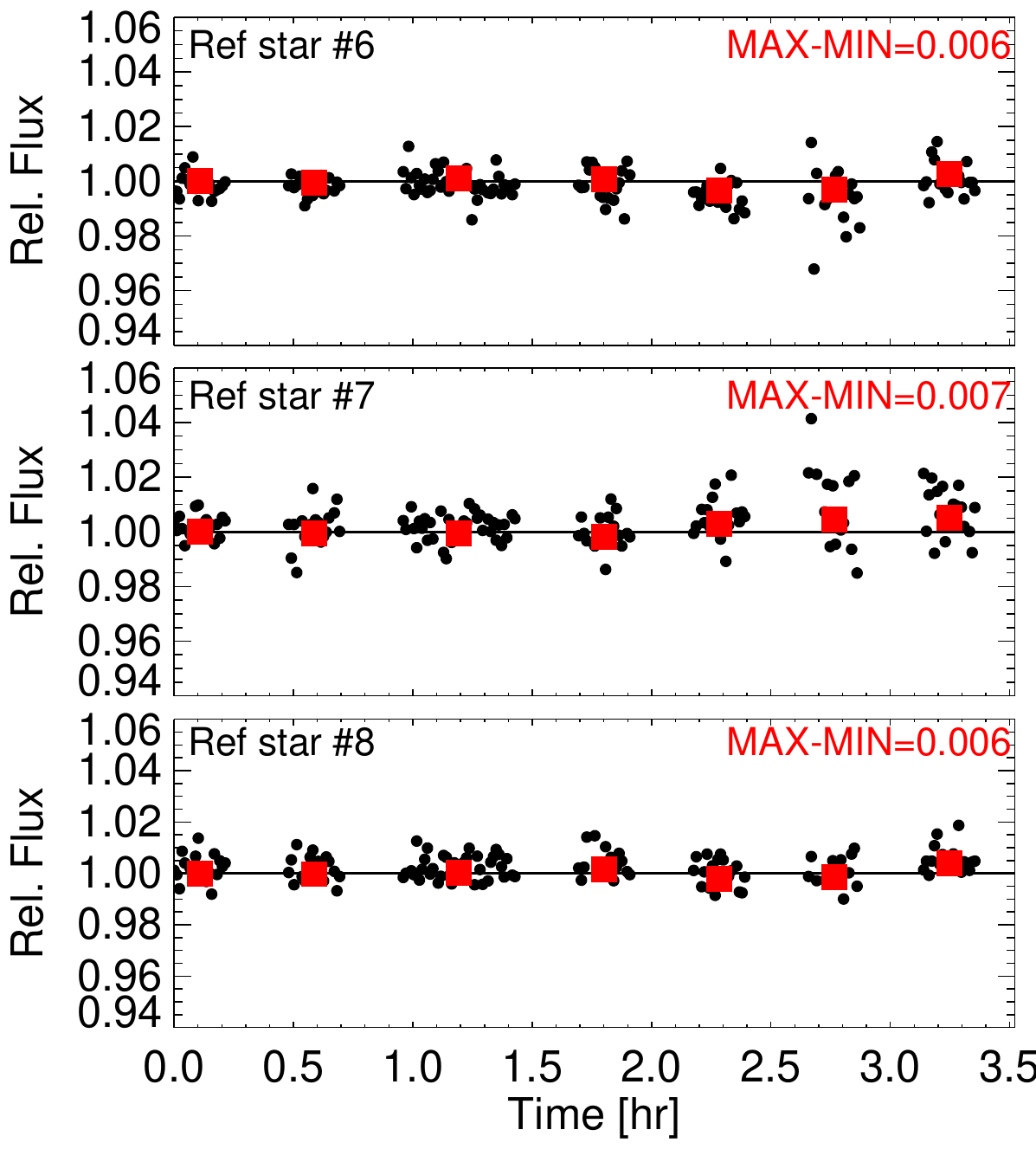}} &
\multirow{1}{*}[0.15in]{\hspace{-0.25in}\includegraphics[width=0.36\hsize]{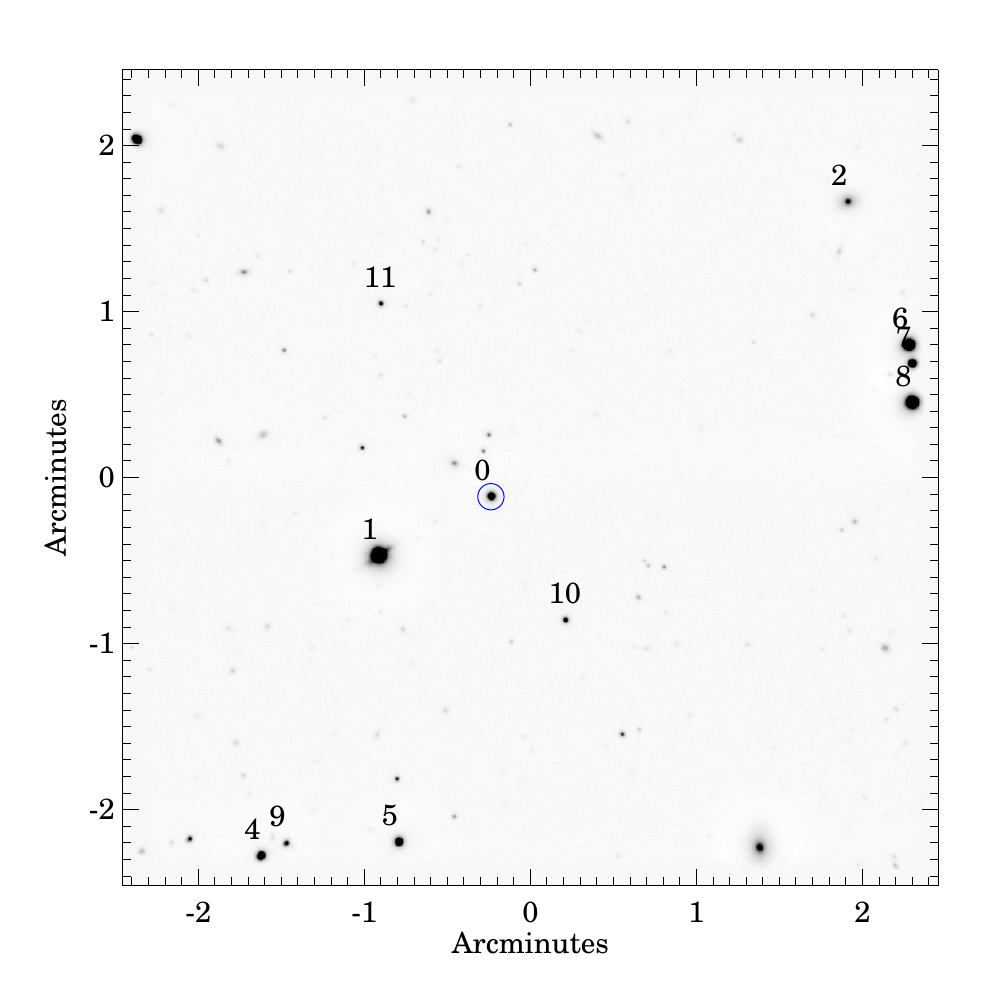}} 
\end{tabular}
\vspace{2.3 in}\caption{Same as figure \ref{fig:lc1}, but for the L0 dwarf 2M0106-59.  W14 report a target amplitude of 4.3$\pm$1.2\% for this time series. \label{fig:lc4}}
\end{figure*}

\vspace{-3mm}\begin{figure*}[ht!]
\begin{tabular}{ccc}
\multirow{2}{*}[0in]{\includegraphics[width=0.3\hsize]{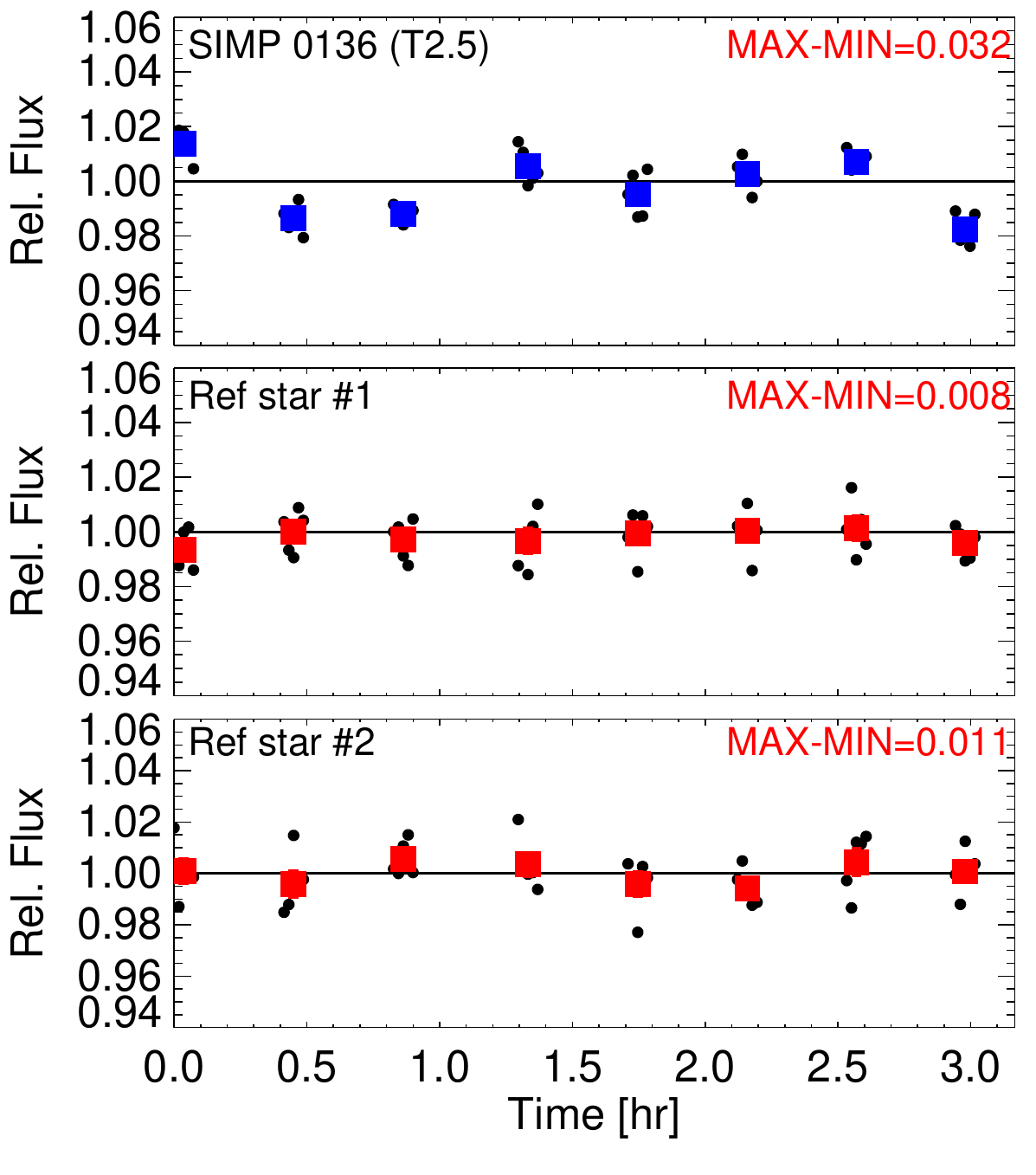}} &
\multirow{2}{*}[0in]{\hspace{-0.2 in} \includegraphics[width=0.3\hsize]{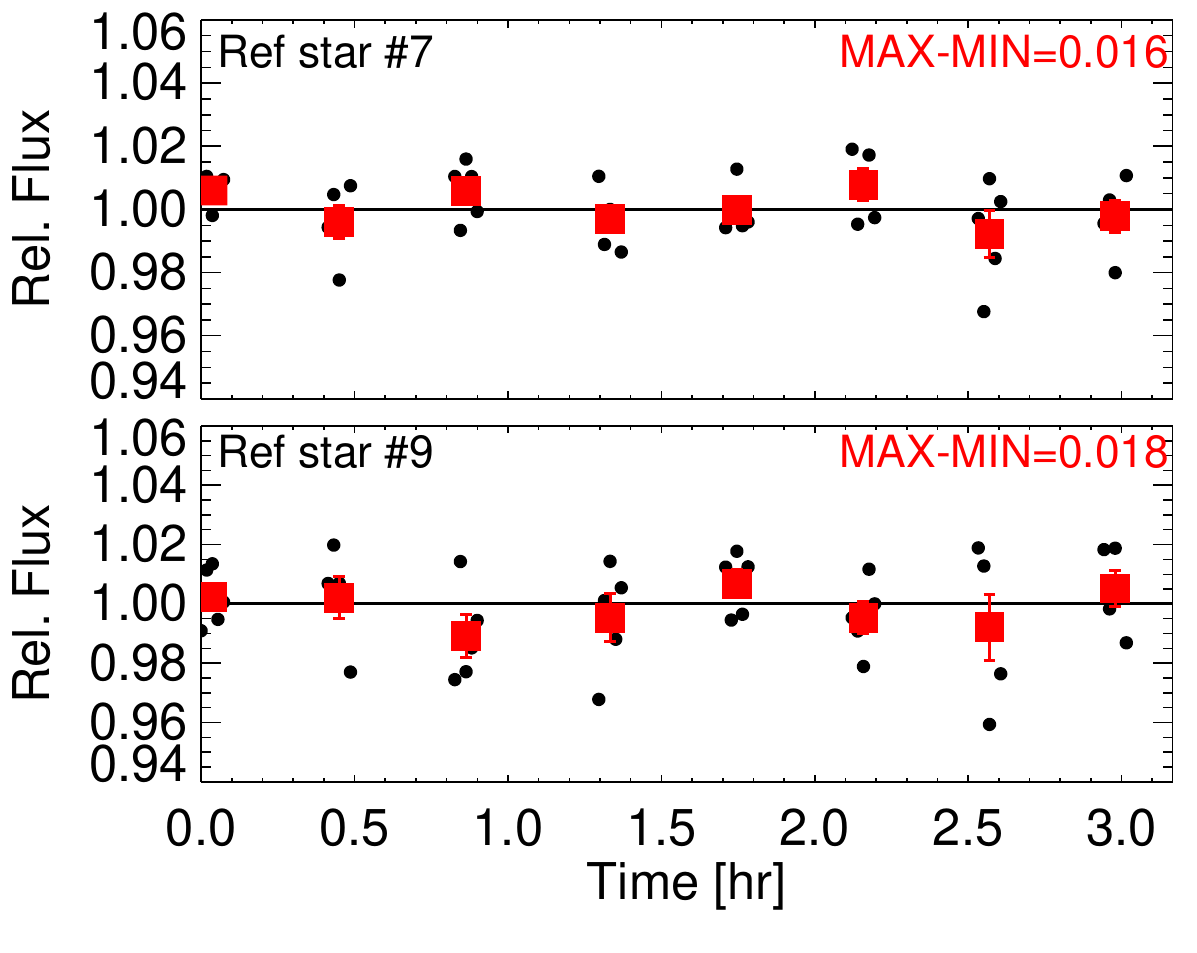}} &
\multirow{1}{*}[0.15in]{\hspace{-0.25in}\includegraphics[width=0.36\hsize]{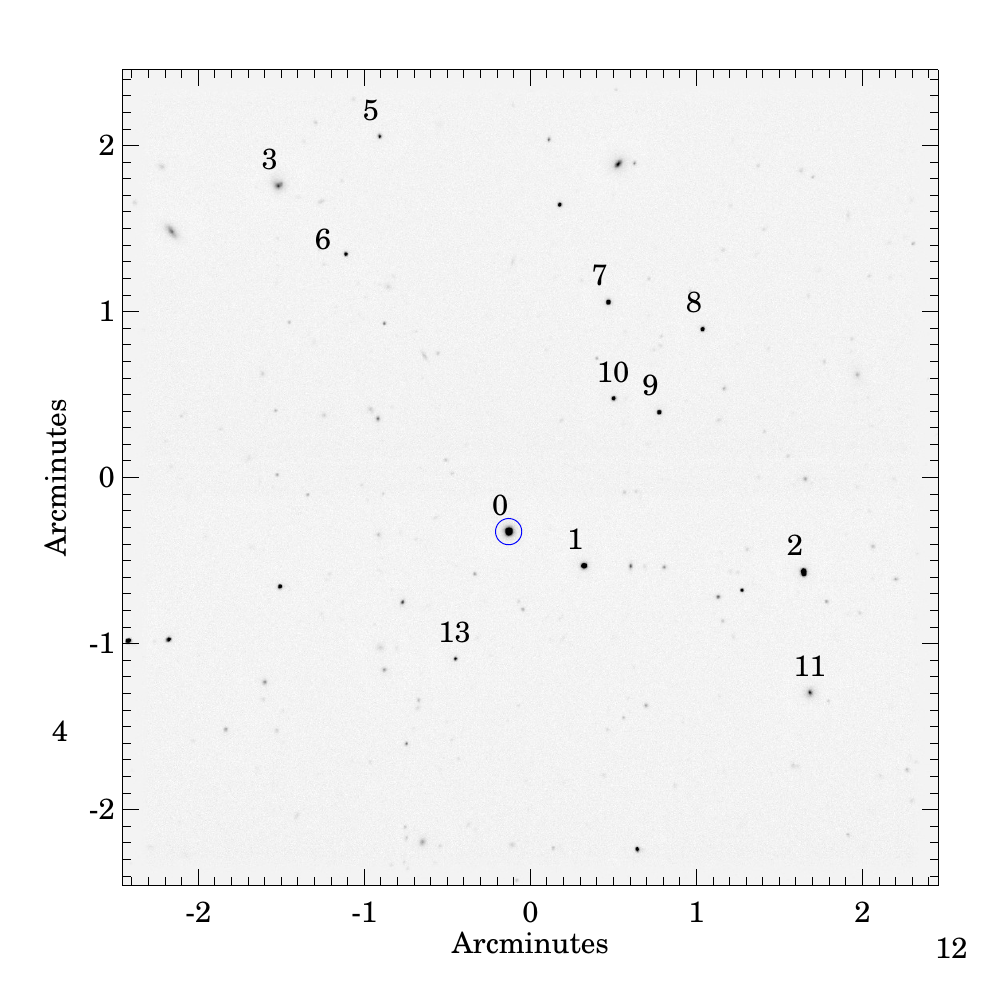}} 
\end{tabular}
\vspace{2.3 in}\caption{Same as figure \ref{fig:lc1} but for the T2.5 dwarf SIMP0136.  W14 report a target amplitude of 3.0$\pm$0.6\% for this time series. \label{fig:lc5}}
\end{figure*}

\begin{figure*}[ht!]
\begin{tabular}{ccc}
\multirow{2}{*}[0in]{\includegraphics[width=0.3\hsize]{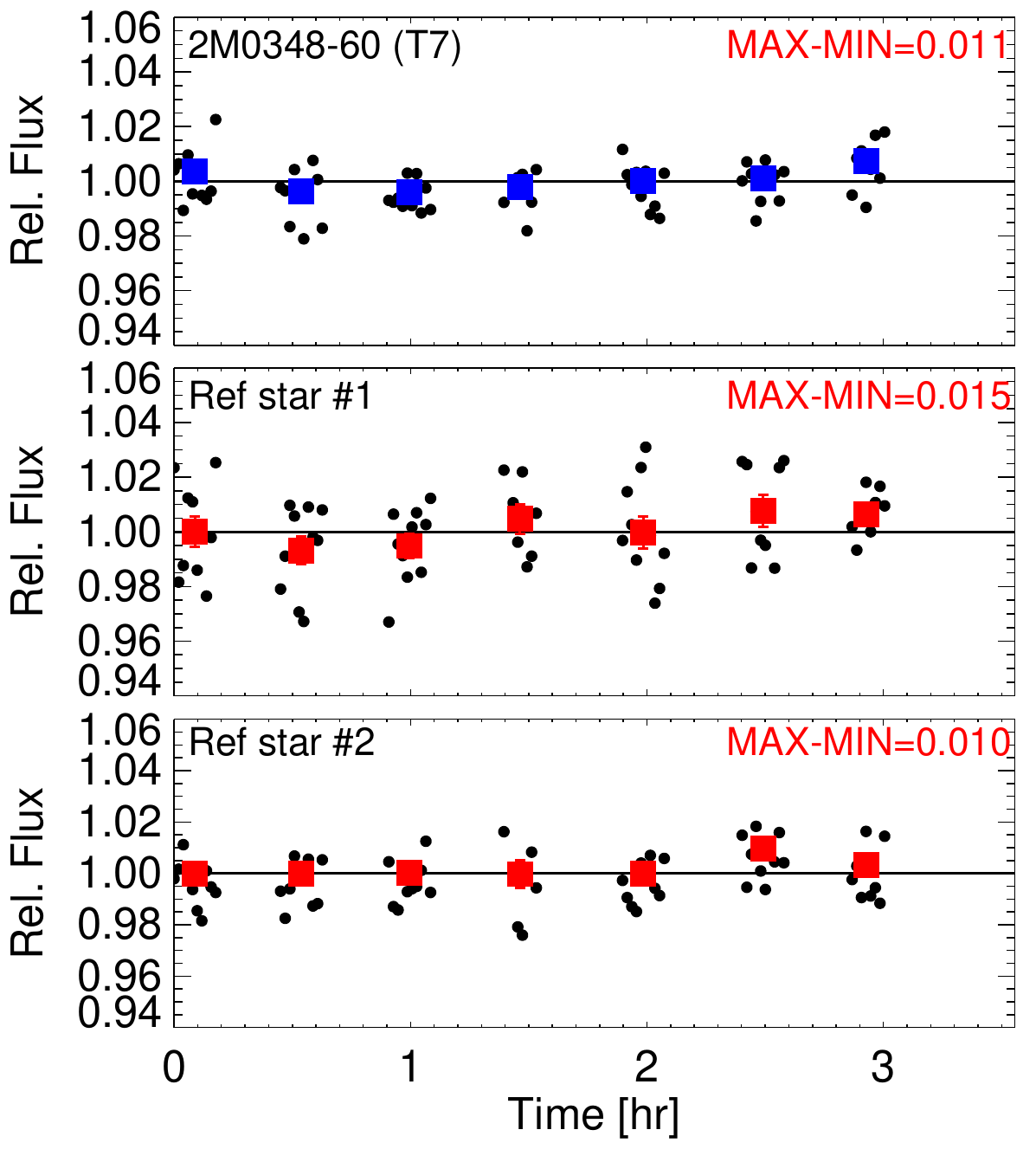}} &
\multirow{2}{*}[0in]{\hspace{-0.2 in} \includegraphics[width=0.3\hsize]{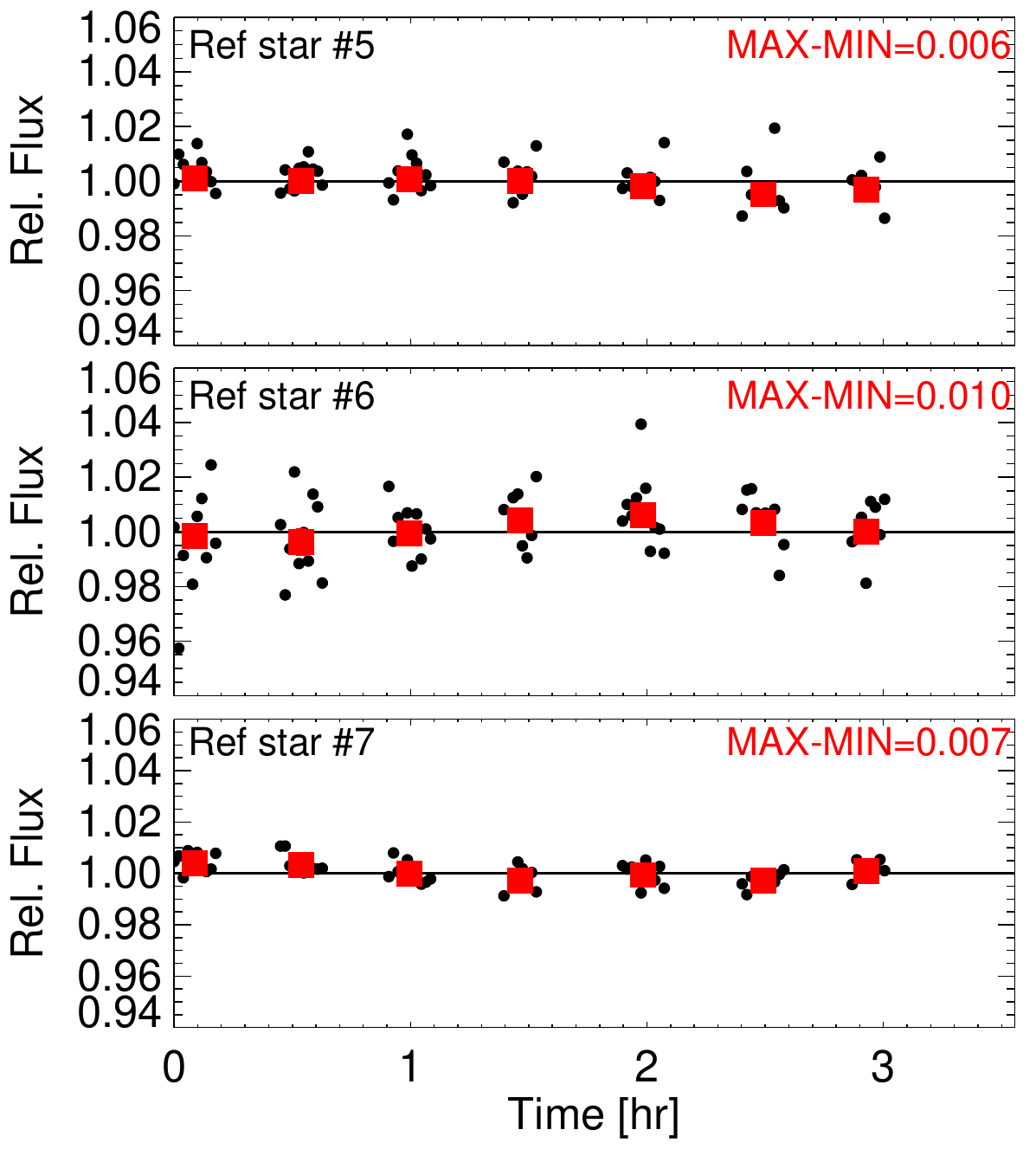}} &
\multirow{1}{*}[0.15in]{\hspace{-0.25in}\includegraphics[width=0.36\hsize]{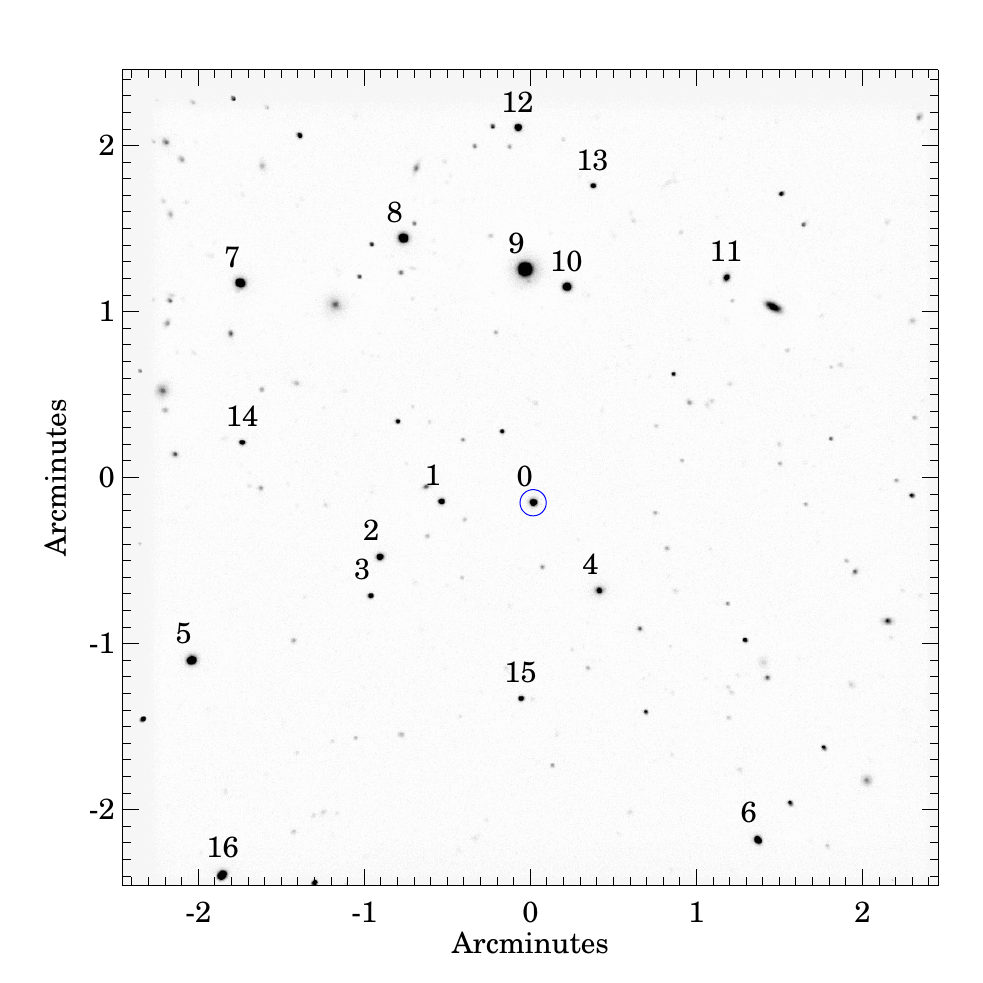}} 
\end{tabular}
\vspace{2.3 in}\caption{Same as figure \ref{fig:lc1} but for the T7 dwarf 0348$-$60. W14 report a target amplitude of 2.4$\pm$0.5\% for this time series. \label{fig:lc6}}
\end{figure*}

\clearpage

\begin{figure*}[ht!]
\begin{tabular}{ccc}
\multirow{2}{*}[0in]{\includegraphics[width=0.3\hsize]{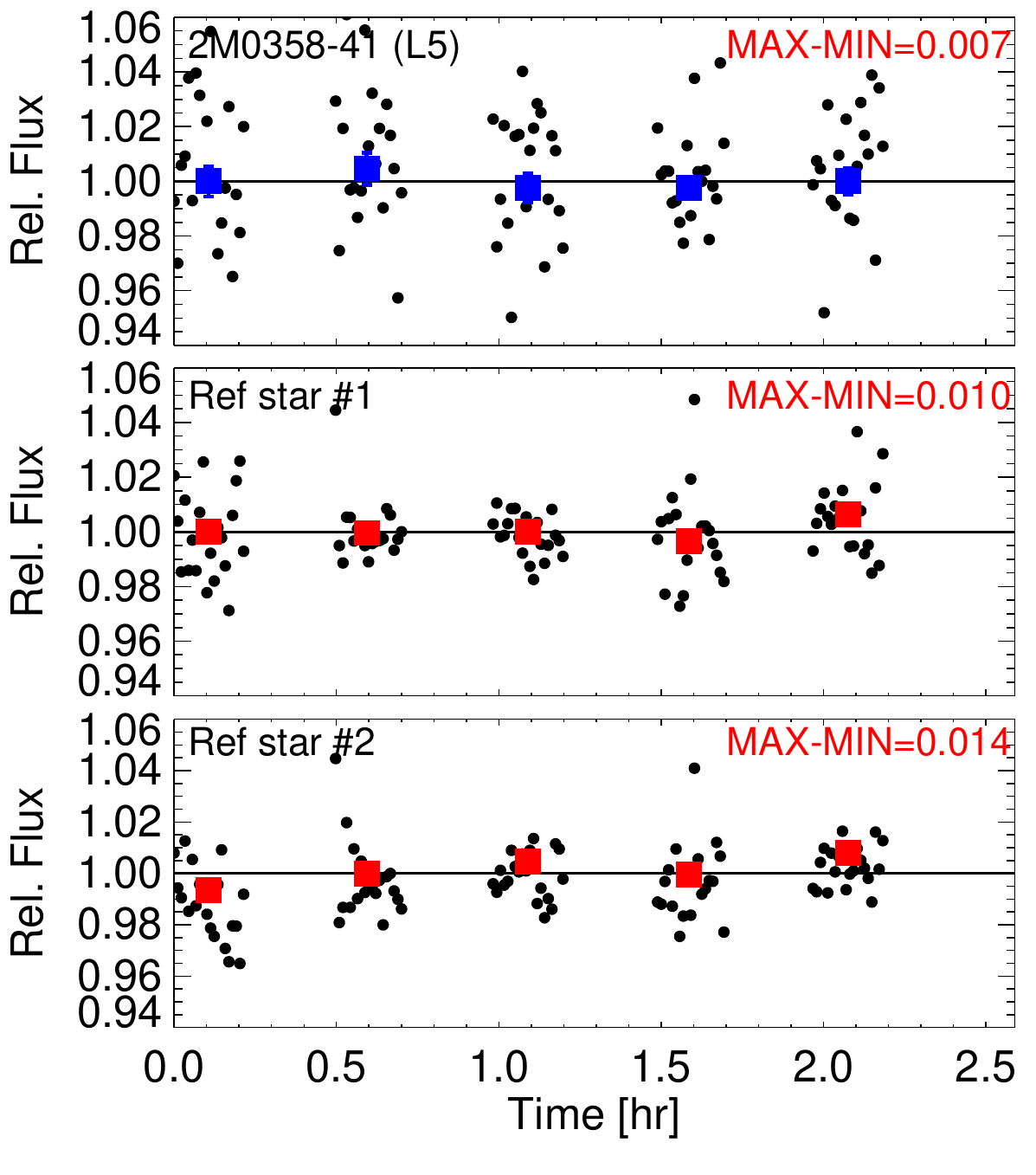}} &
\multirow{2}{*}[0in]{\hspace{-0.2 in} \includegraphics[width=0.3\hsize]{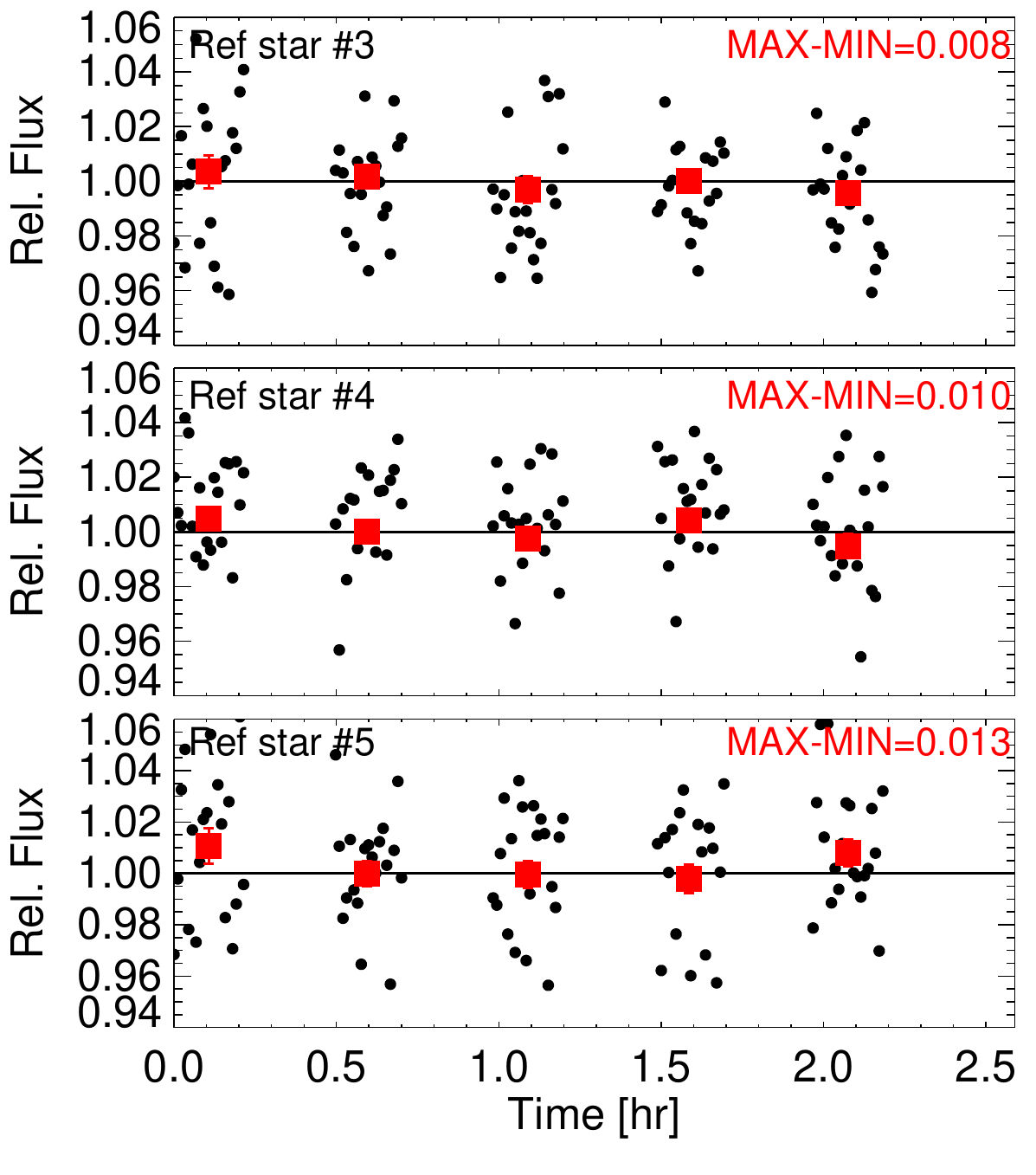}} &
\multirow{1}{*}[0.15in]{\hspace{-0.25in}\includegraphics[width=0.36\hsize]{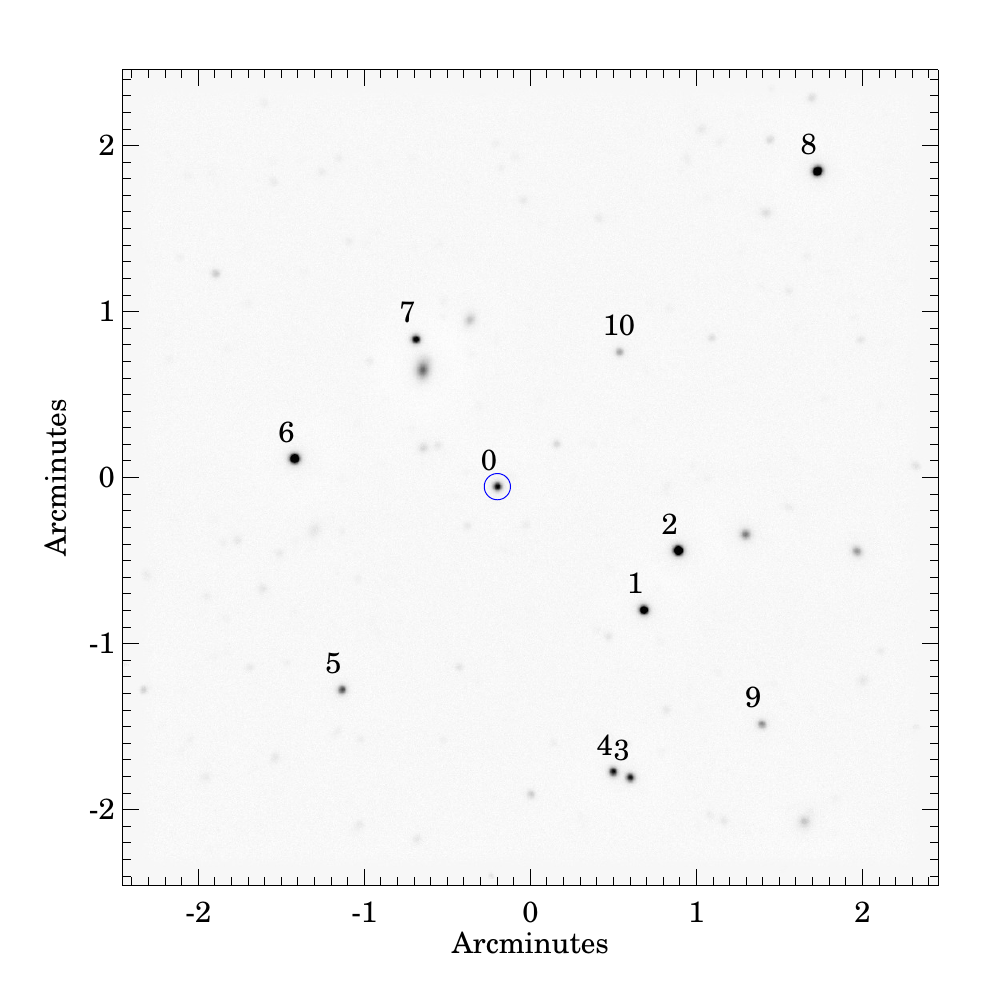}} 
\end{tabular}
\vspace{2.3 in}\caption{Same as figure \ref{fig:lc1} but for the L5 dwarf 0358$-$41. W14 report a target amplitude of 4.8$\pm$1.2\% for this time series. \label{fig:lc7}}
\end{figure*}

\begin{figure*}[ht!]
\begin{tabular}{ccc}
\multirow{2}{*}[0in]{\includegraphics[width=0.3\hsize]{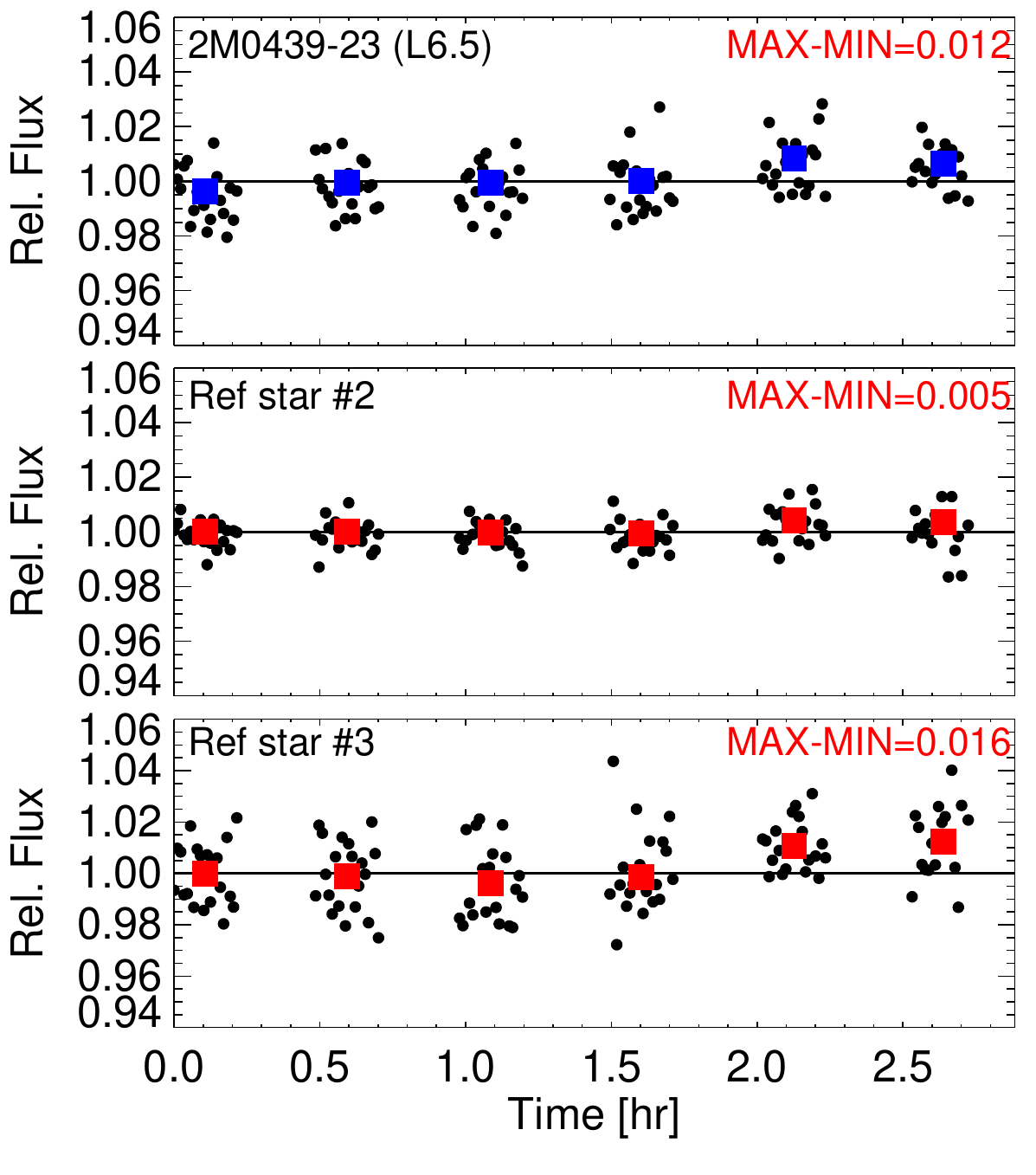}} &
\multirow{2}{*}[0in]{\hspace{-0.2 in} \includegraphics[width=0.3\hsize]{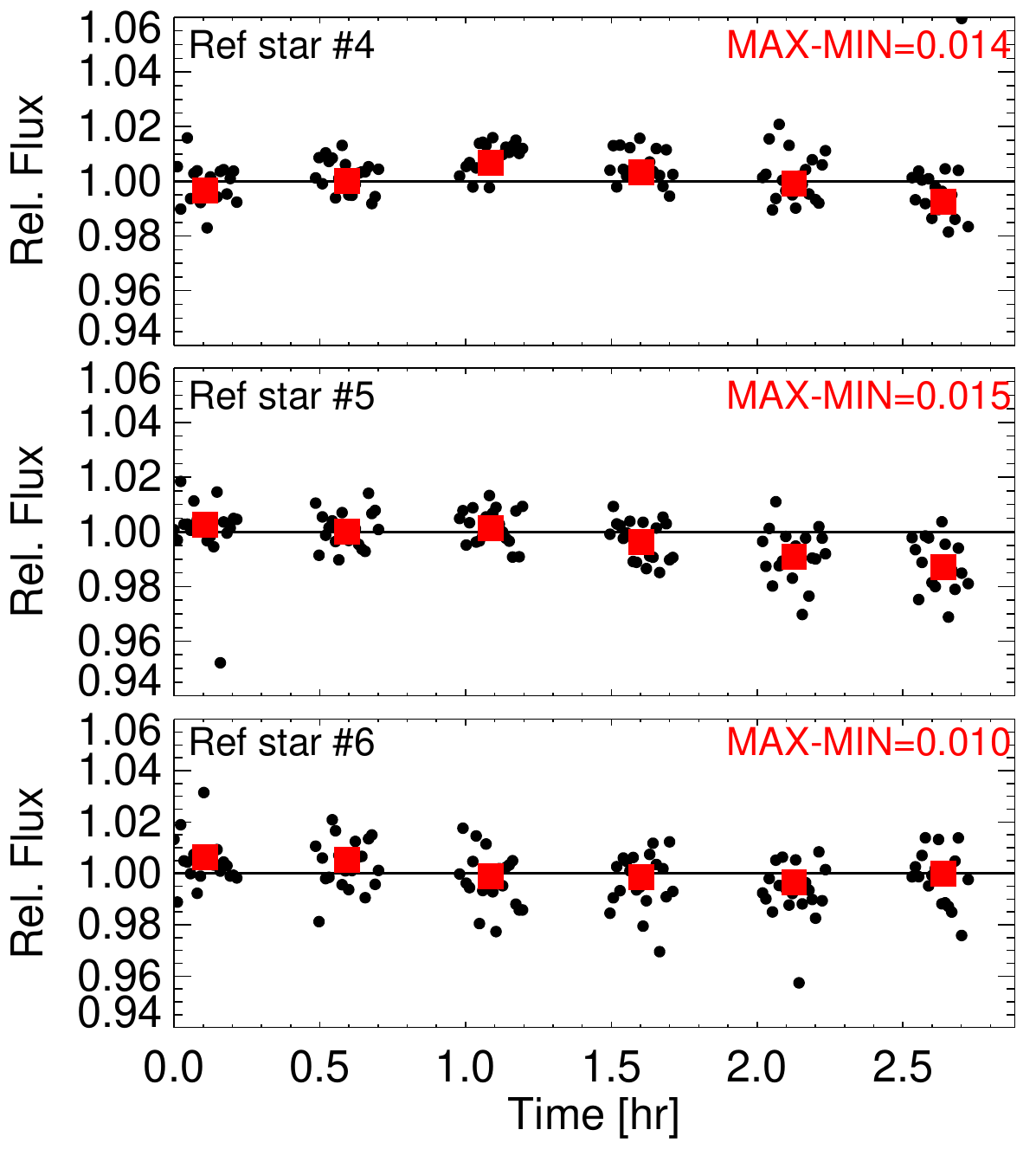}} &
\multirow{1}{*}[0.15in]{\hspace{-0.25in}\includegraphics[width=0.36\hsize]{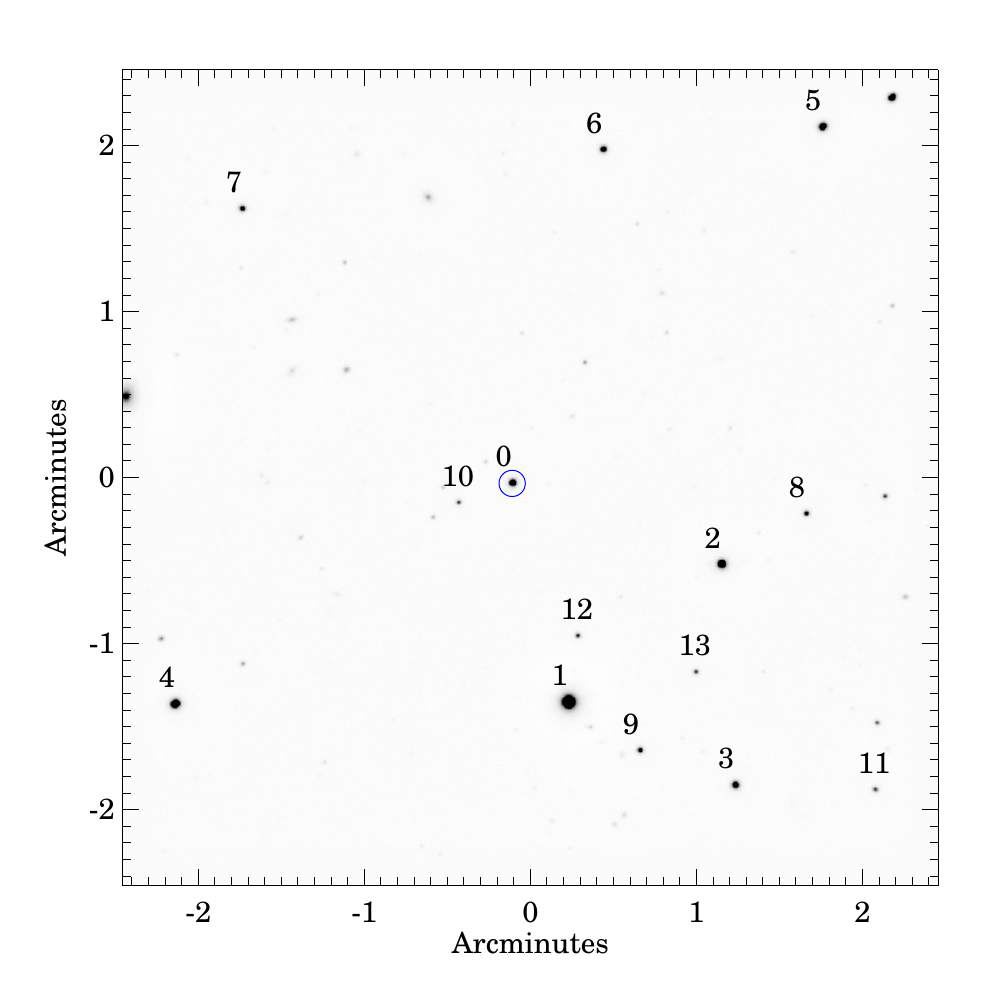}} 
\end{tabular}
\vspace{2.3 in}\caption{Same as figure \ref{fig:lc1} but for the L6.5 dwarf 2M0439-23.  W14 report a target amplitude of 2.6$\pm$0.5\% for this time series.  \label{fig:lc7}}
\end{figure*}

\begin{figure*}[ht!]
\begin{tabular}{ccc}
\multirow{2}{*}[0in]{\includegraphics[width=0.3\hsize]{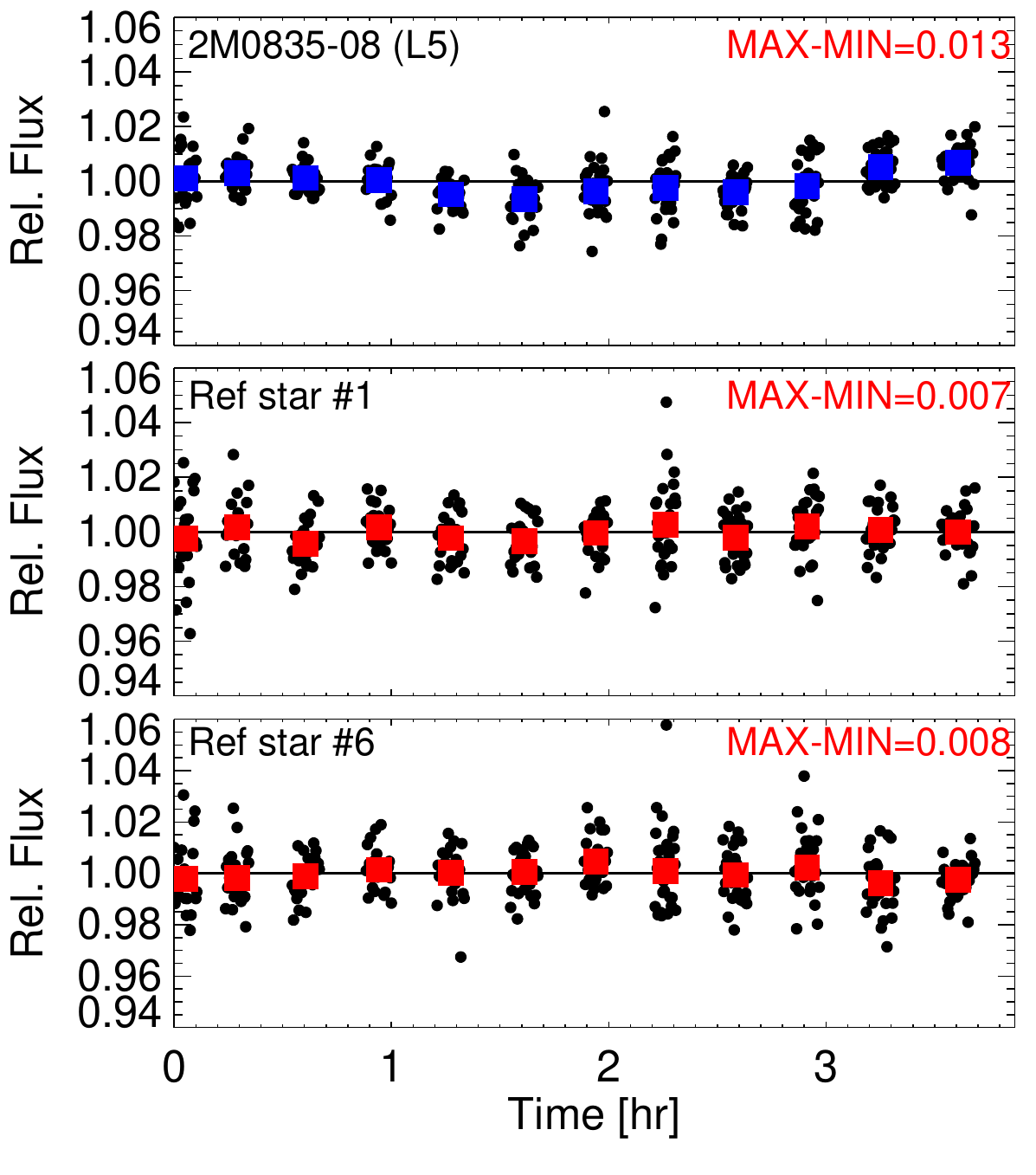}} &
\multirow{2}{*}[0in]{\hspace{-0.2 in} \includegraphics[width=0.3\hsize]{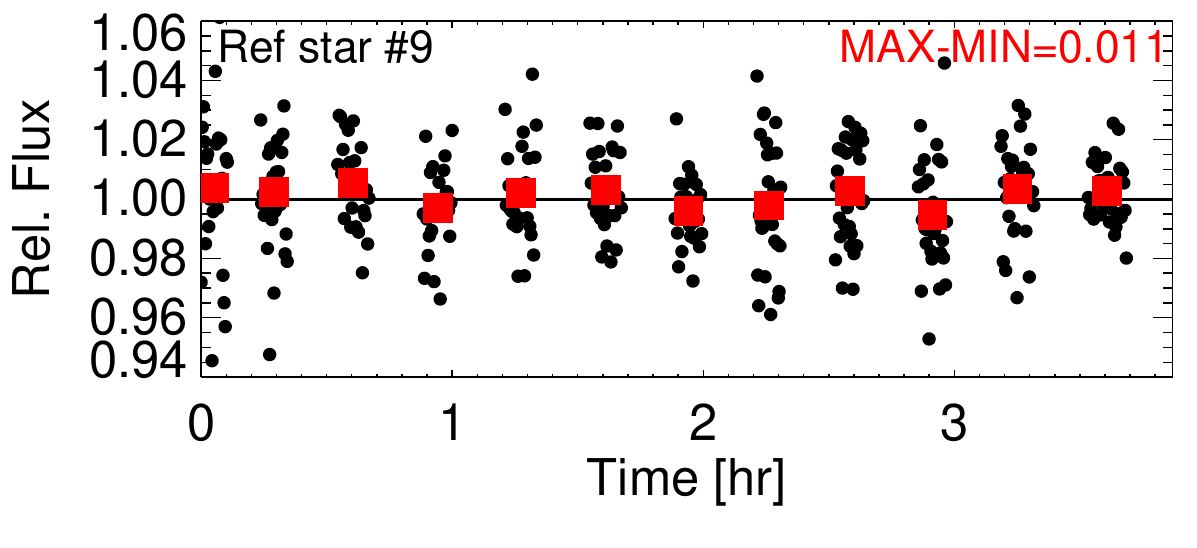}} &
\multirow{1}{*}[0.15in]{\hspace{-0.25in}\includegraphics[width=0.36\hsize]{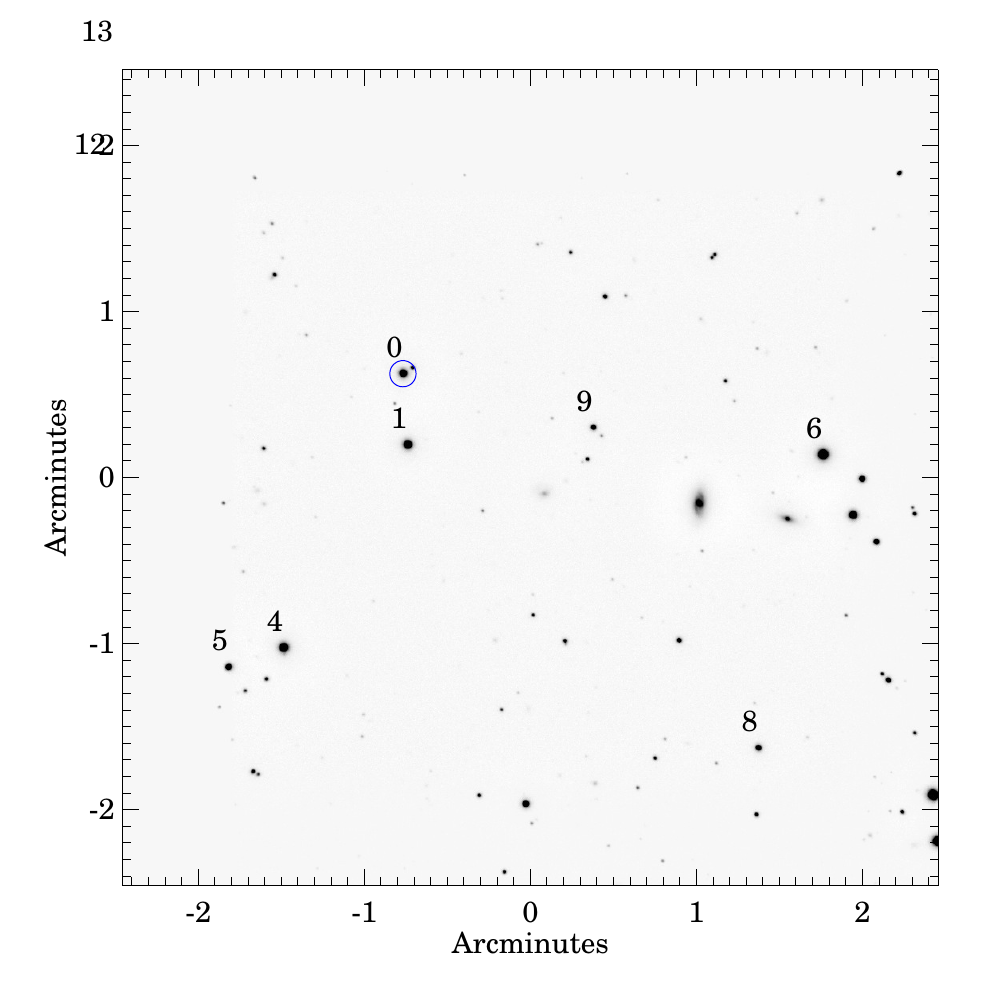}} 
\end{tabular}
\vspace{2.3 in}\caption{Same as figure \ref{fig:lc1} but for the L5 dwarf 2M0835-08.  W14 report a target amplitude of 1.7$\pm$0.5\% for this time series.  \label{fig:lc8}}
\end{figure*}

\clearpage

\begin{figure*}[ht!]
\begin{tabular}{ccc}
\multirow{2}{*}[0in]{\includegraphics[width=0.3\hsize]{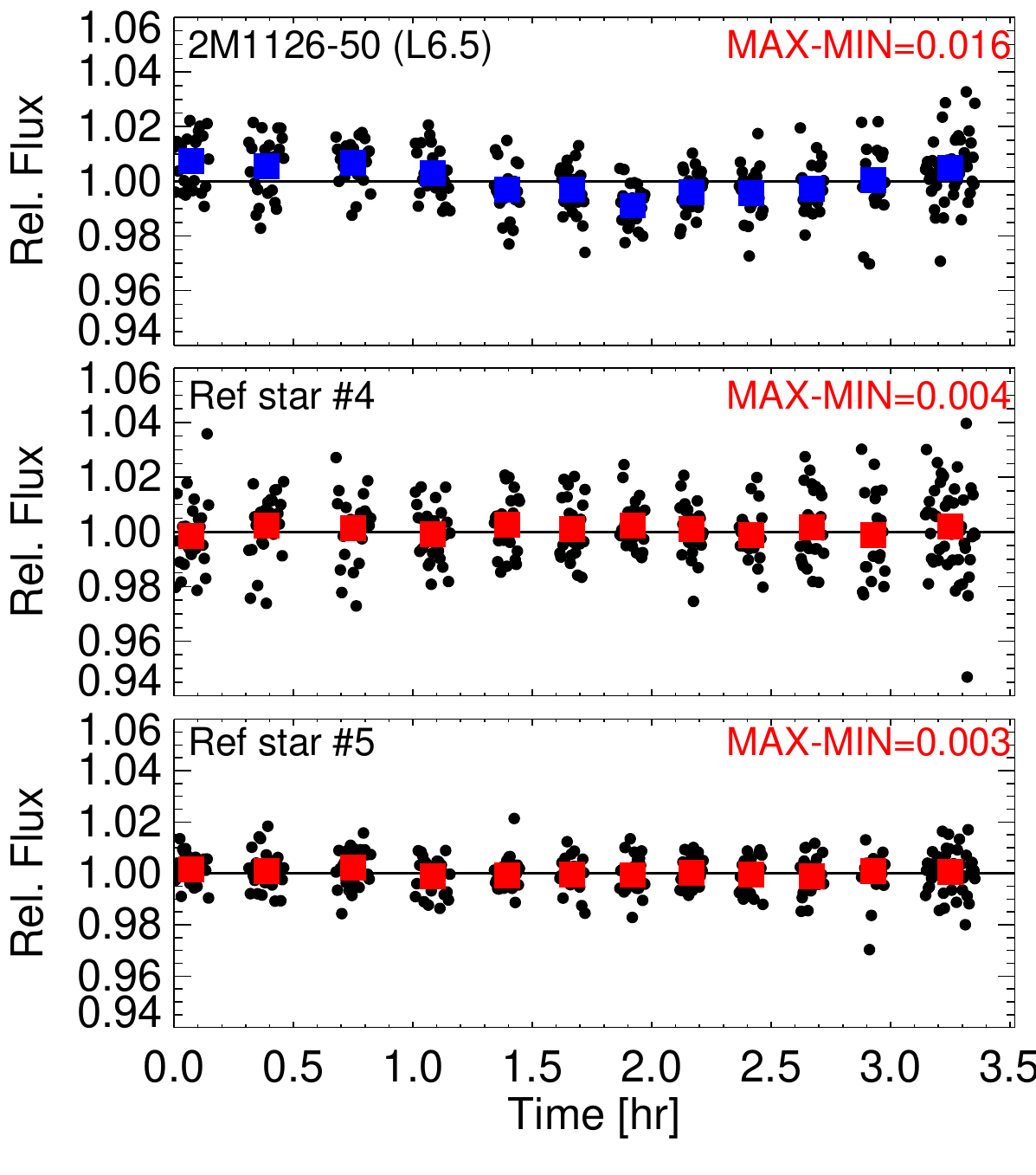}} &
\multirow{2}{*}[0in]{\hspace{-0.2 in} \includegraphics[width=0.3\hsize]{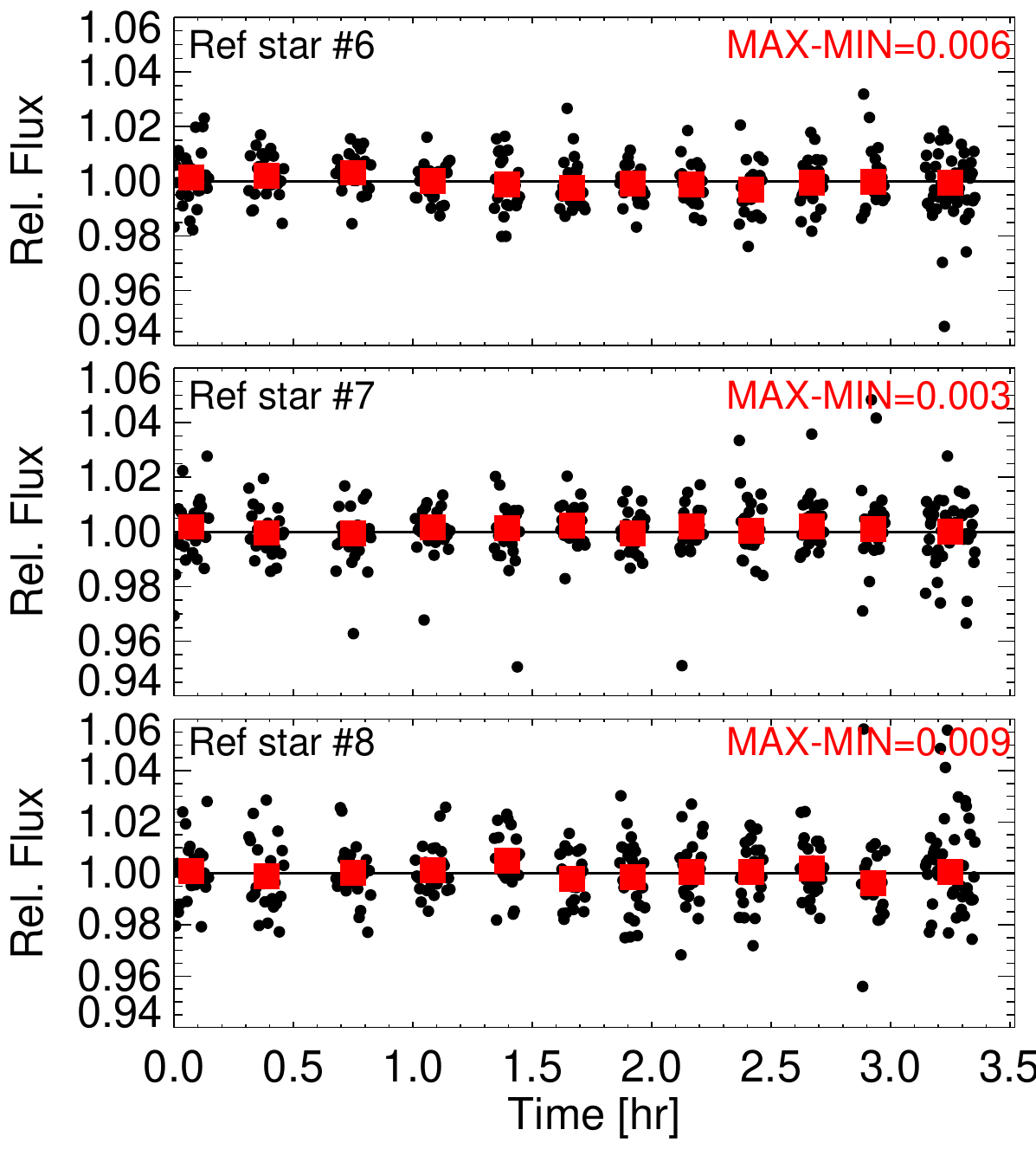}} &
\multirow{1}{*}[0.15in]{\hspace{-0.25in}\includegraphics[width=0.36\hsize]{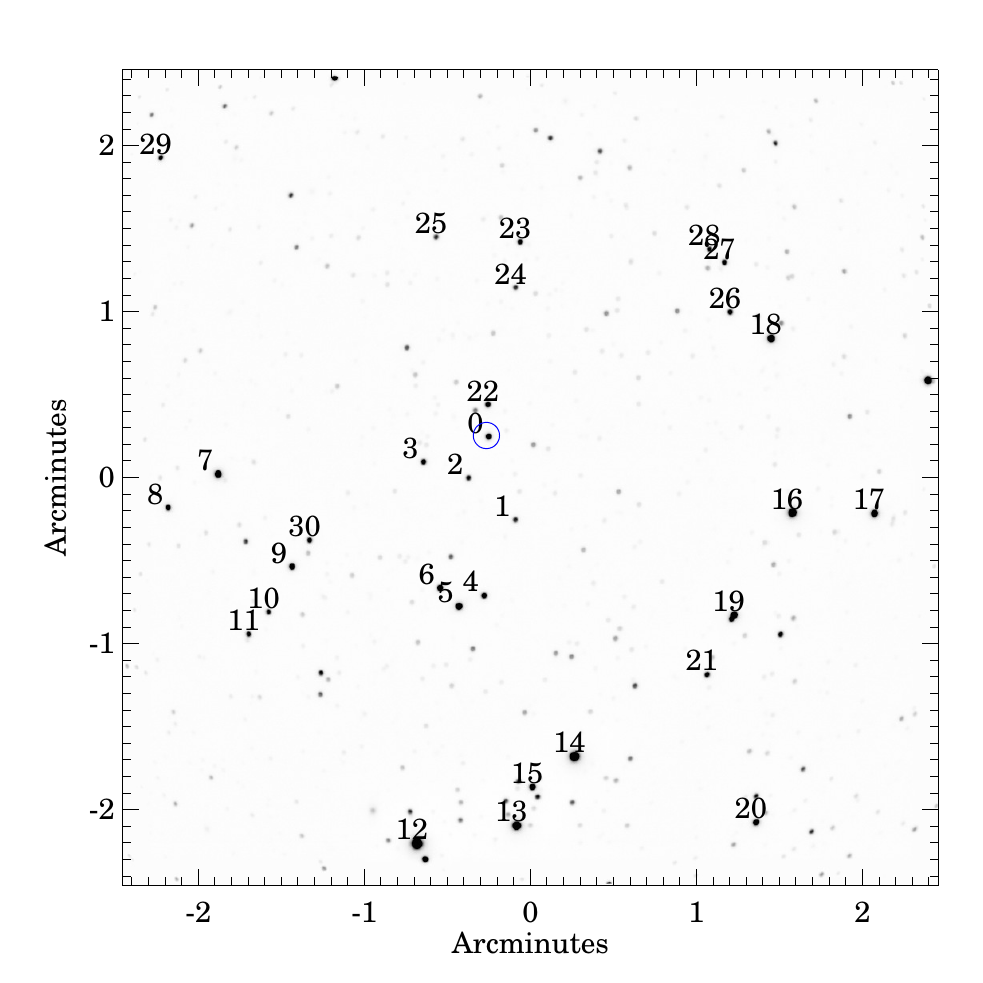}} 
\end{tabular}
\vspace{2.3 in}\caption{Same as figure \ref{fig:lc1} but for the L6.5 dwarf 2M1126-50.  W14 report a target amplitude of 3.2$\pm$0.7\% for this time series.  \label{fig:lc9}}
\end{figure*}

\begin{figure*}[ht!]
\begin{tabular}{ccc}
\multirow{2}{*}[0in]{\includegraphics[width=0.3\hsize]{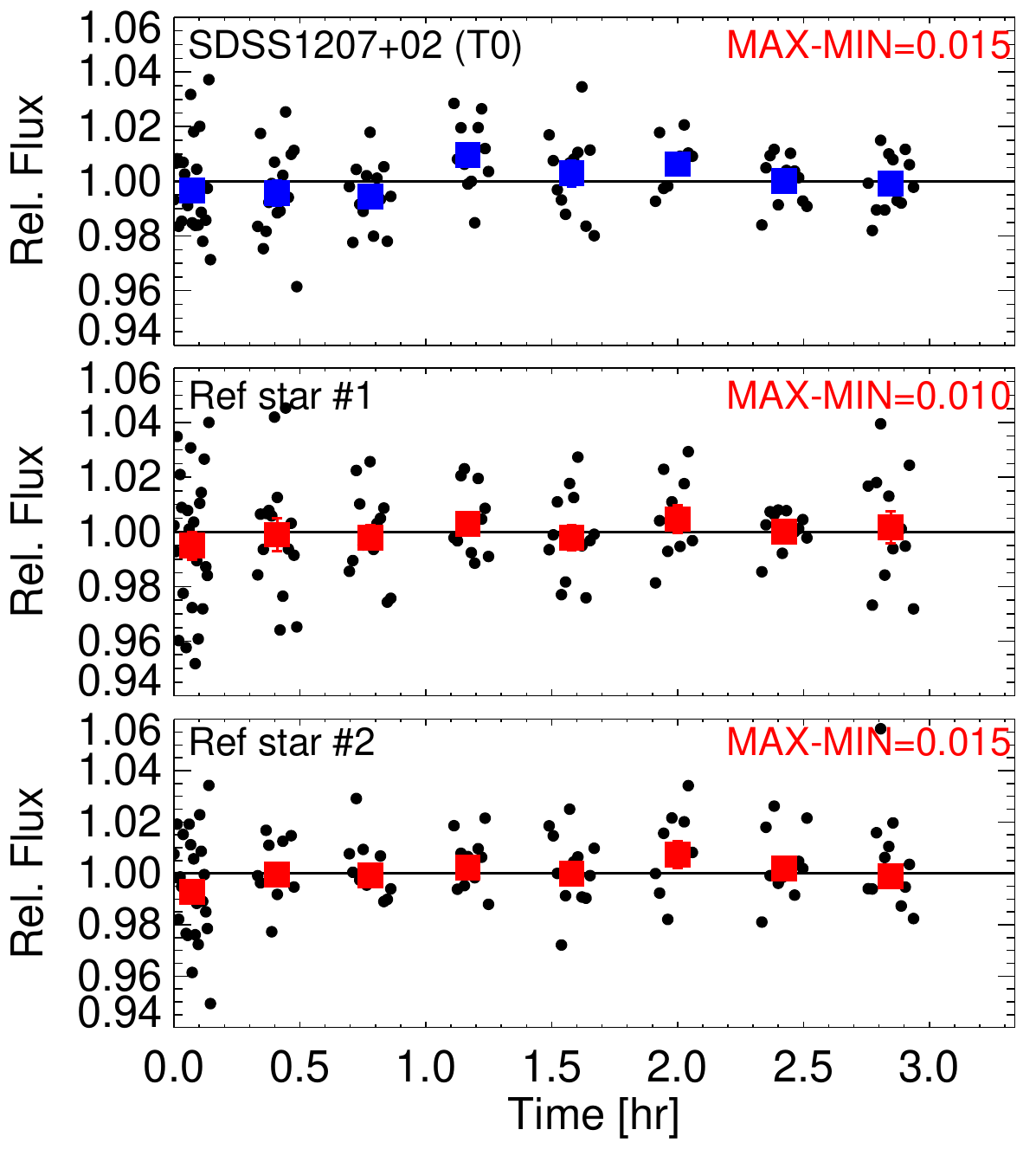}} &
\multirow{2}{*}[0in]{\hspace{-0.2 in} \includegraphics[width=0.3\hsize]{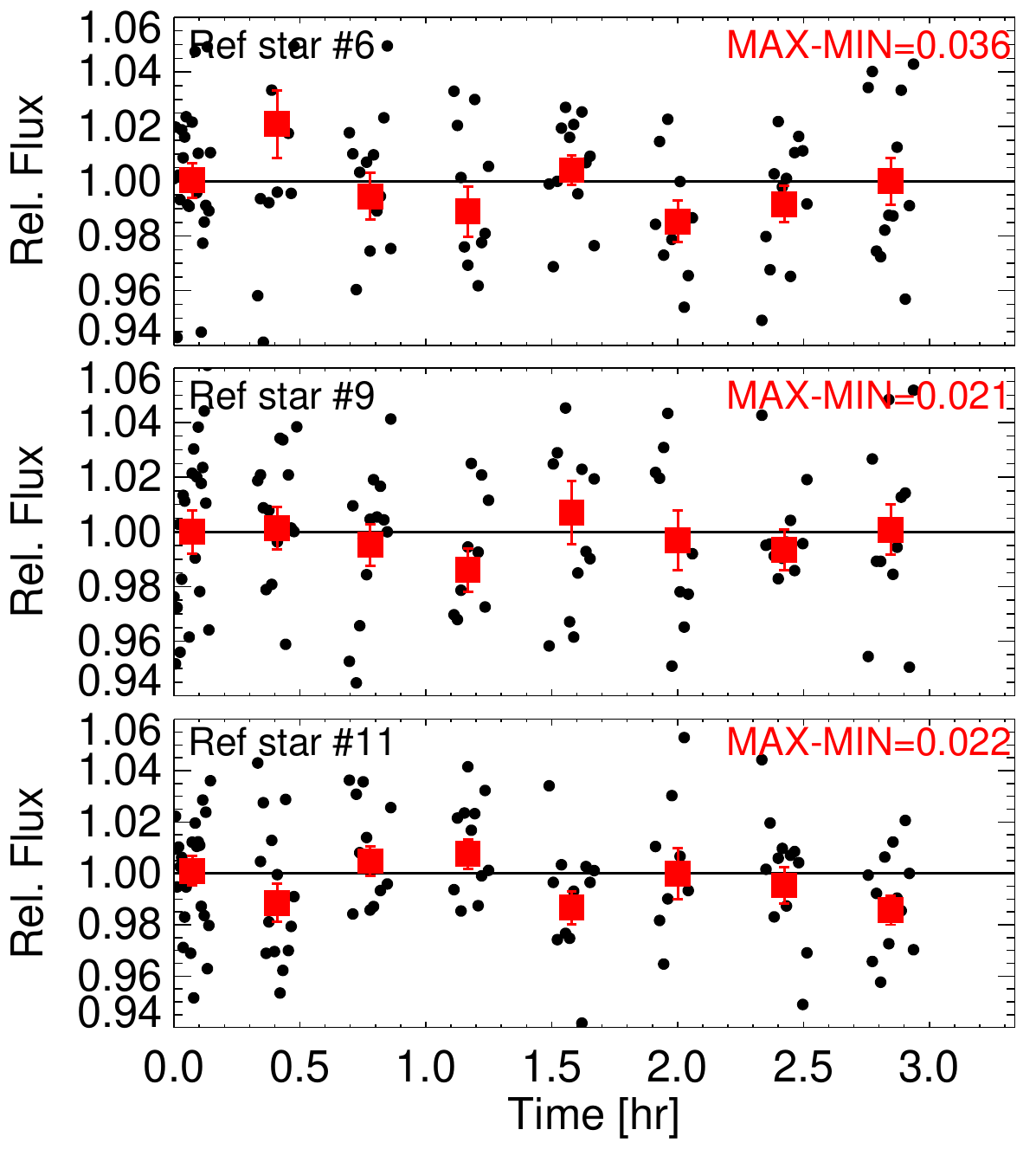}} &
\multirow{1}{*}[0.15in]{\hspace{-0.25in}\includegraphics[width=0.36\hsize]{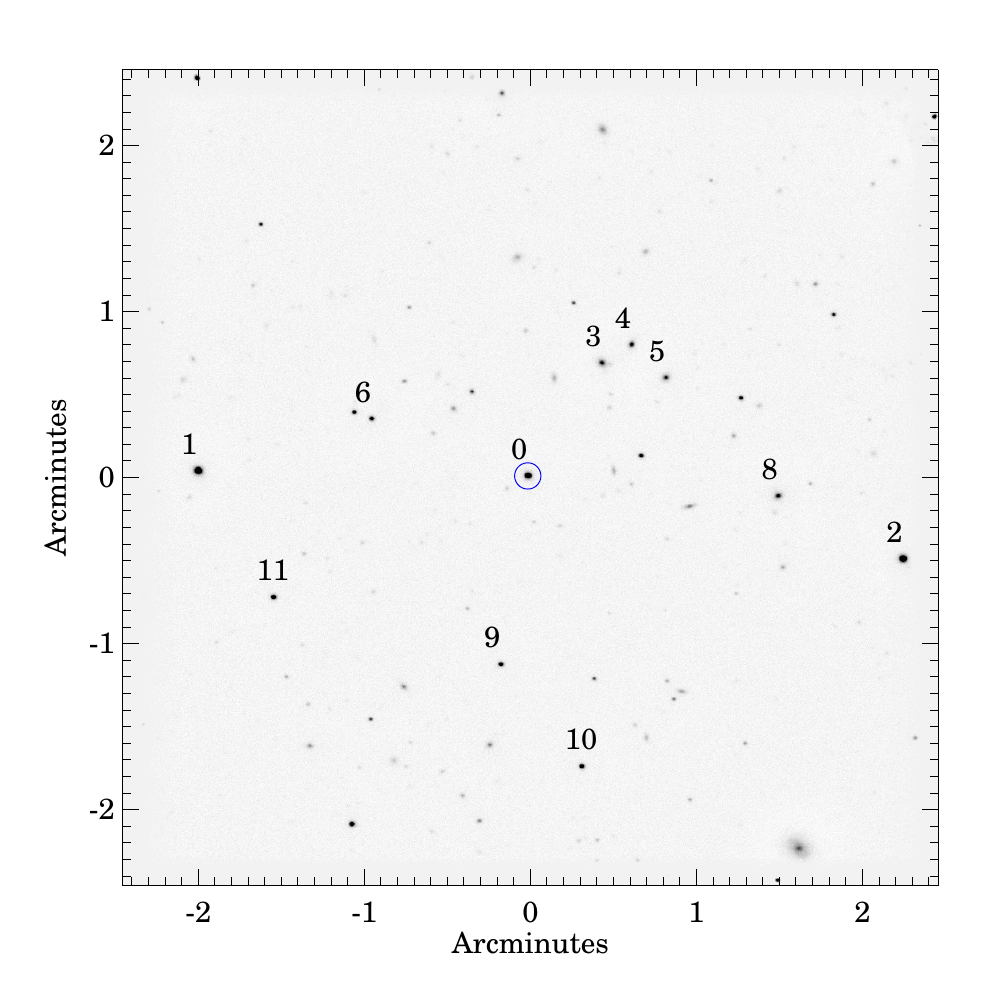}} 
\end{tabular}
\vspace{2.3 in}\caption{Same as figure \ref{fig:lc1} but for the T0 dwarf 2M1207$+$02.  W14 report a target amplitude of 5.2$\pm$1.1\% for this time series.  \label{fig:lc10}}
\end{figure*}

\begin{figure*}[ht!]
\begin{tabular}{ccc}
\multirow{2}{*}[0in]{\includegraphics[width=0.3\hsize]{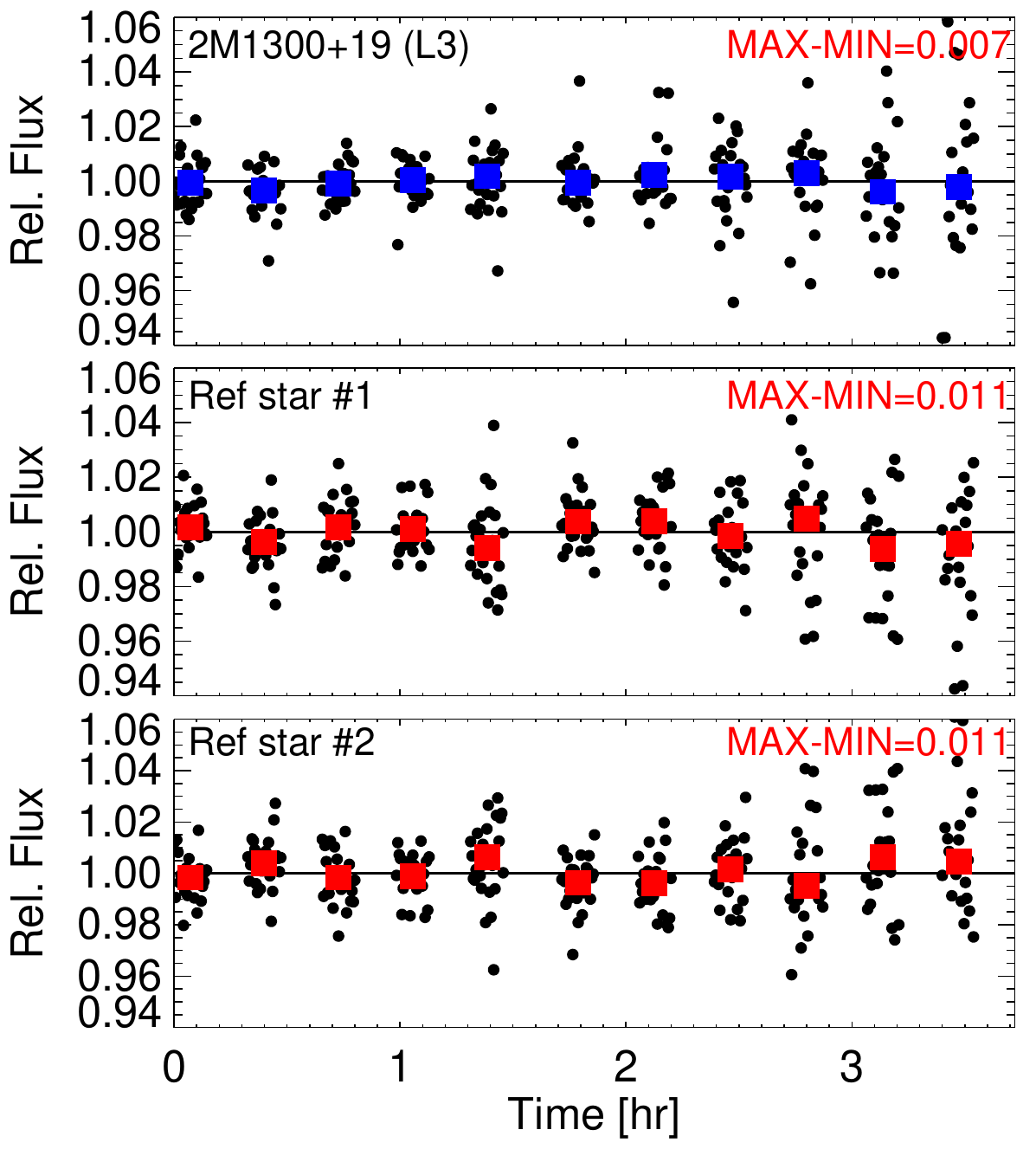}} &
\multirow{1}{*}[0.15in]{\hspace{-0.25in}\includegraphics[width=0.36\hsize]{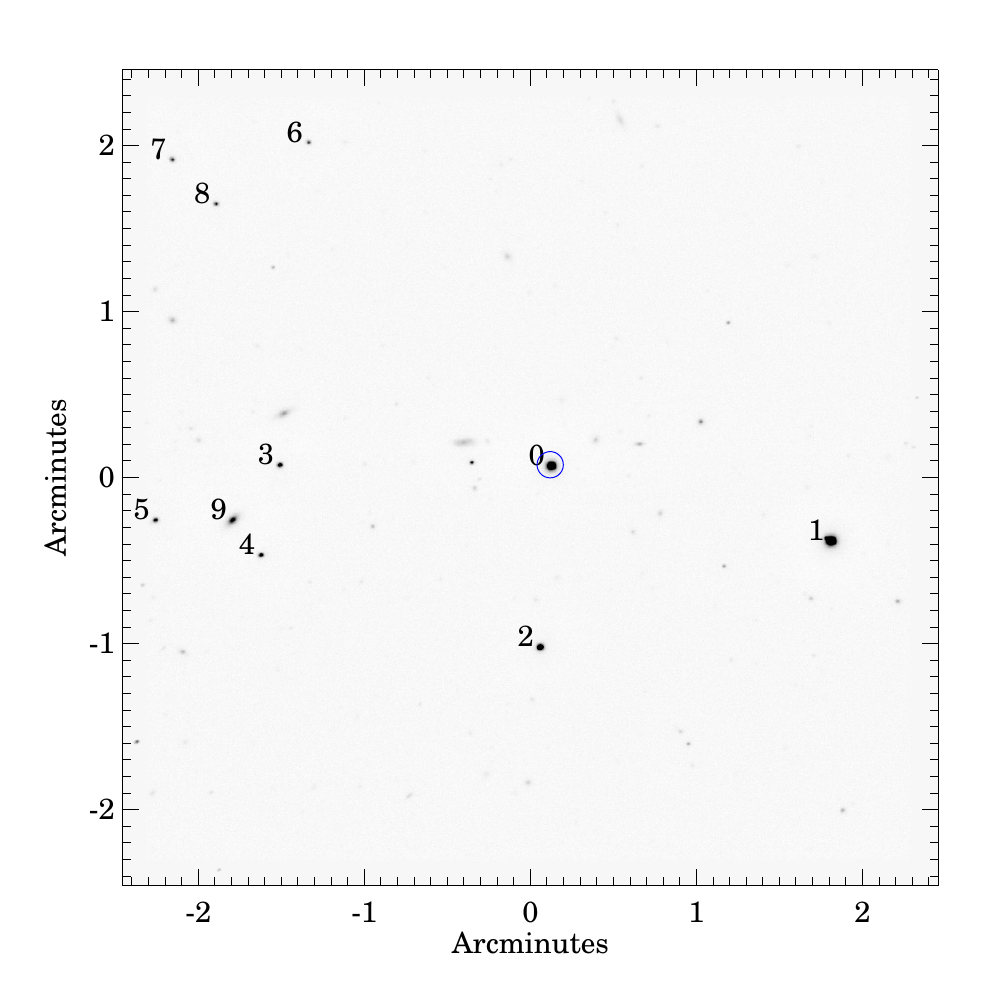}} 
\end{tabular}
\vspace{2.3 in}\caption{Same as figure \ref{fig:lc1} but for the L3 dwarf 2M1300$+$19.  W14 report a target amplitude of 9.6$\pm$0.9\% for this time series.  \label{fig:lc11}}
\end{figure*}

\clearpage

\begin{figure*}[ht!]
\begin{tabular}{ccc}
\multirow{2}{*}[0in]{\includegraphics[width=0.3\hsize]{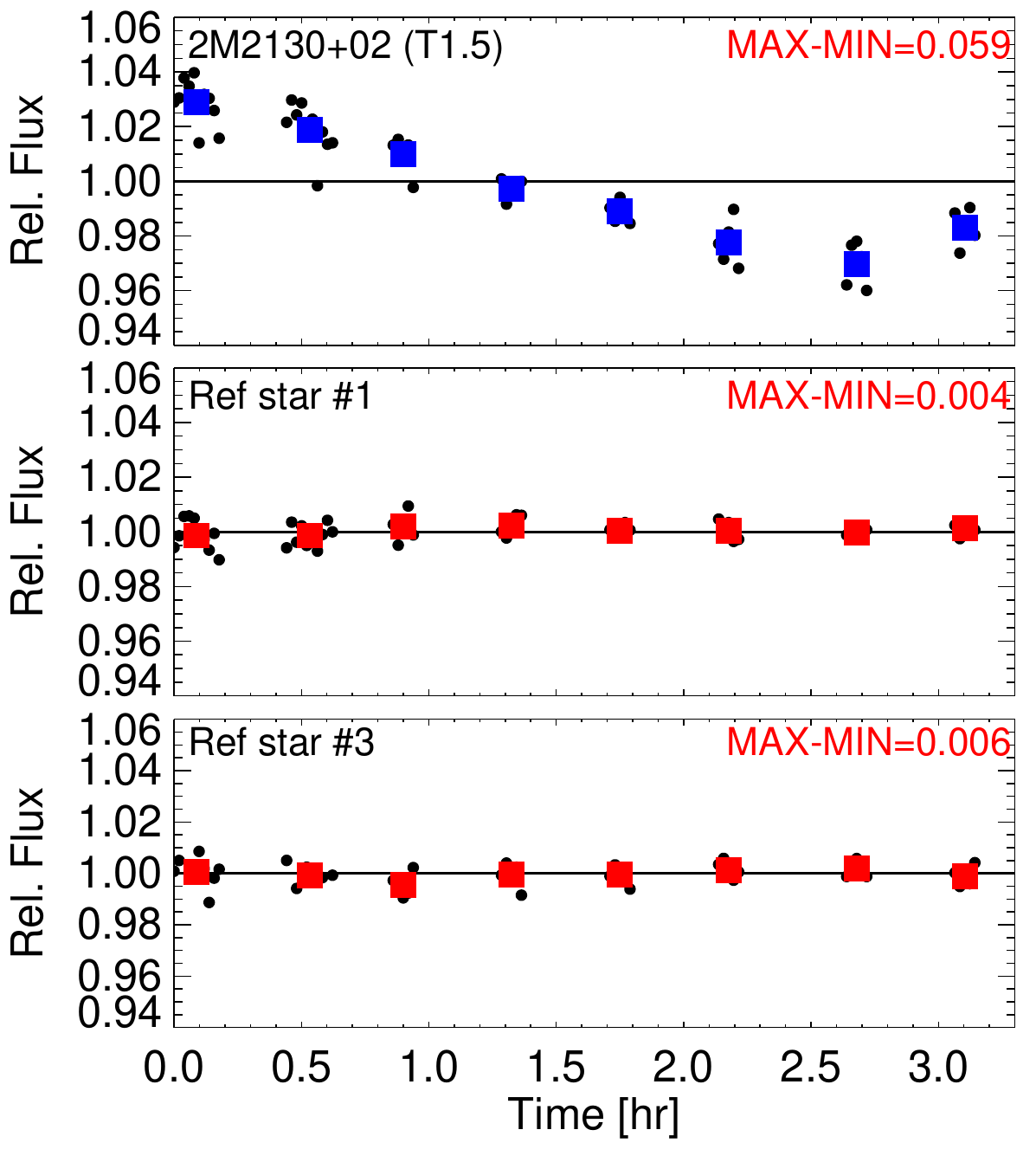}} &
\multirow{2}{*}[0in]{\hspace{-0.2 in} \includegraphics[width=0.3\hsize]{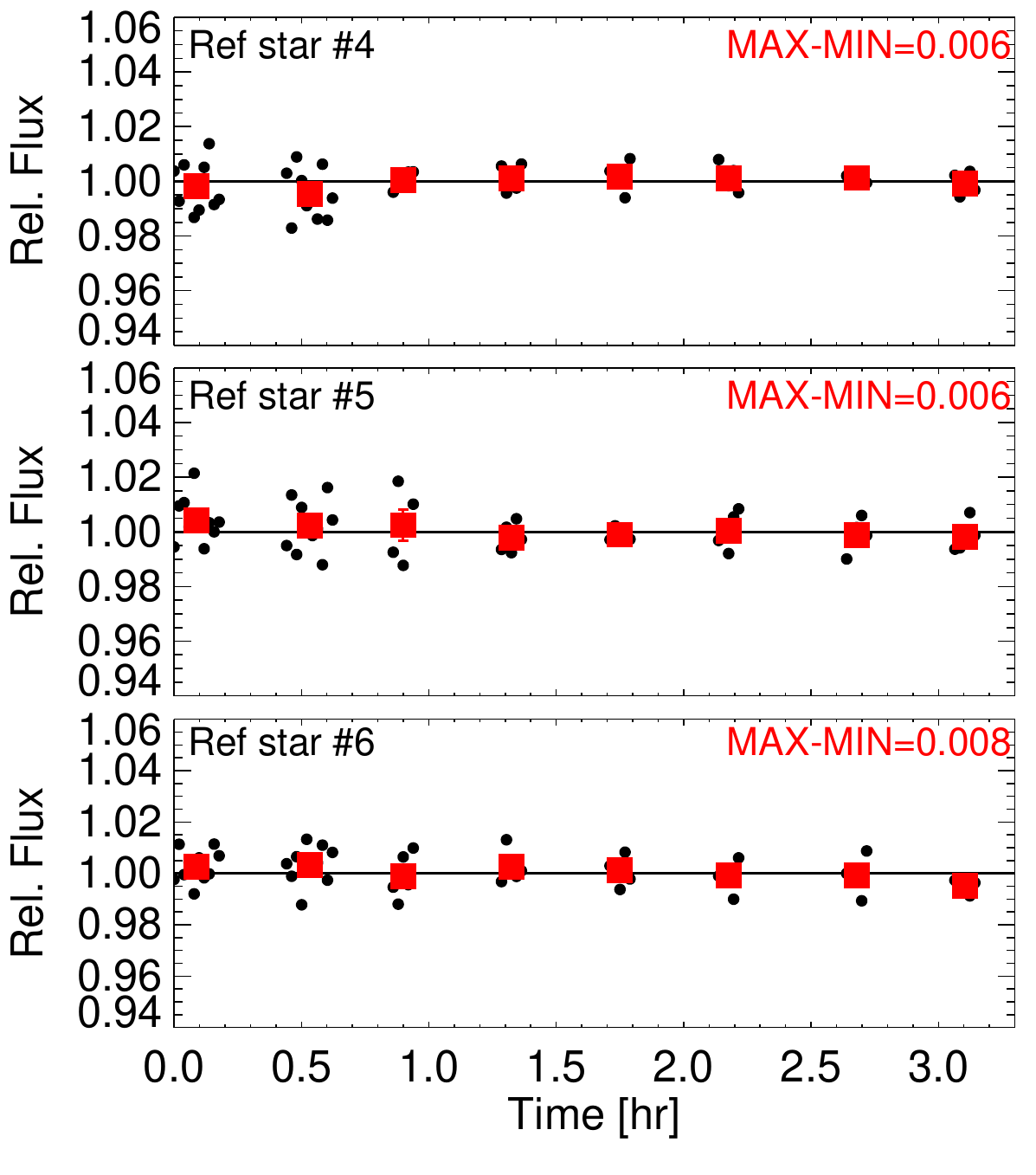}} &
\multirow{1}{*}[0.15in]{\hspace{-0.25in}\includegraphics[width=0.36\hsize]{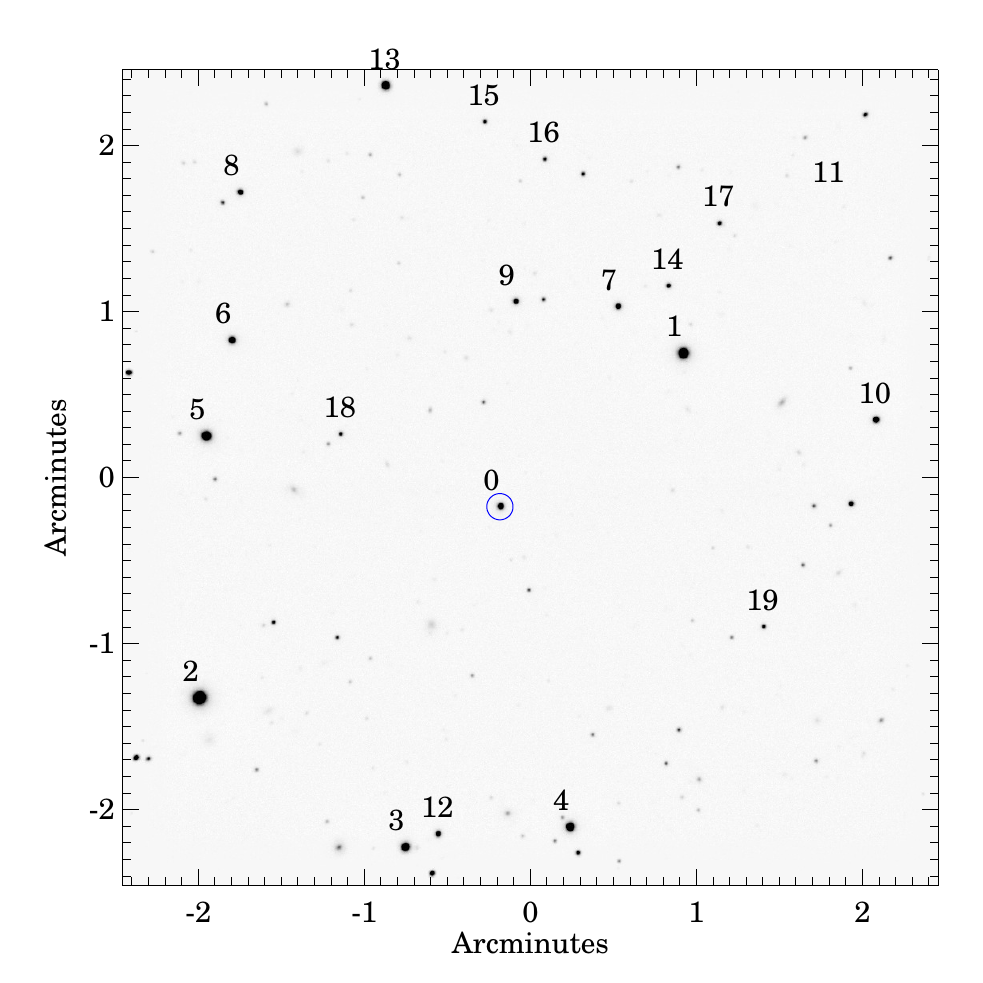}} 
\end{tabular}
\vspace{2.3 in}\caption{Same as figure \ref{fig:lc1} but for the T1.5 dwarf 2M2139$+$02.  W14 report a target amplitude of 4.7$\pm$0.5\% for this time series.  \label{fig:lc12}}
\end{figure*}

\begin{figure*}[ht!]
\begin{tabular}{ccc}
\multirow{2}{*}[0in]{\includegraphics[width=0.3\hsize]{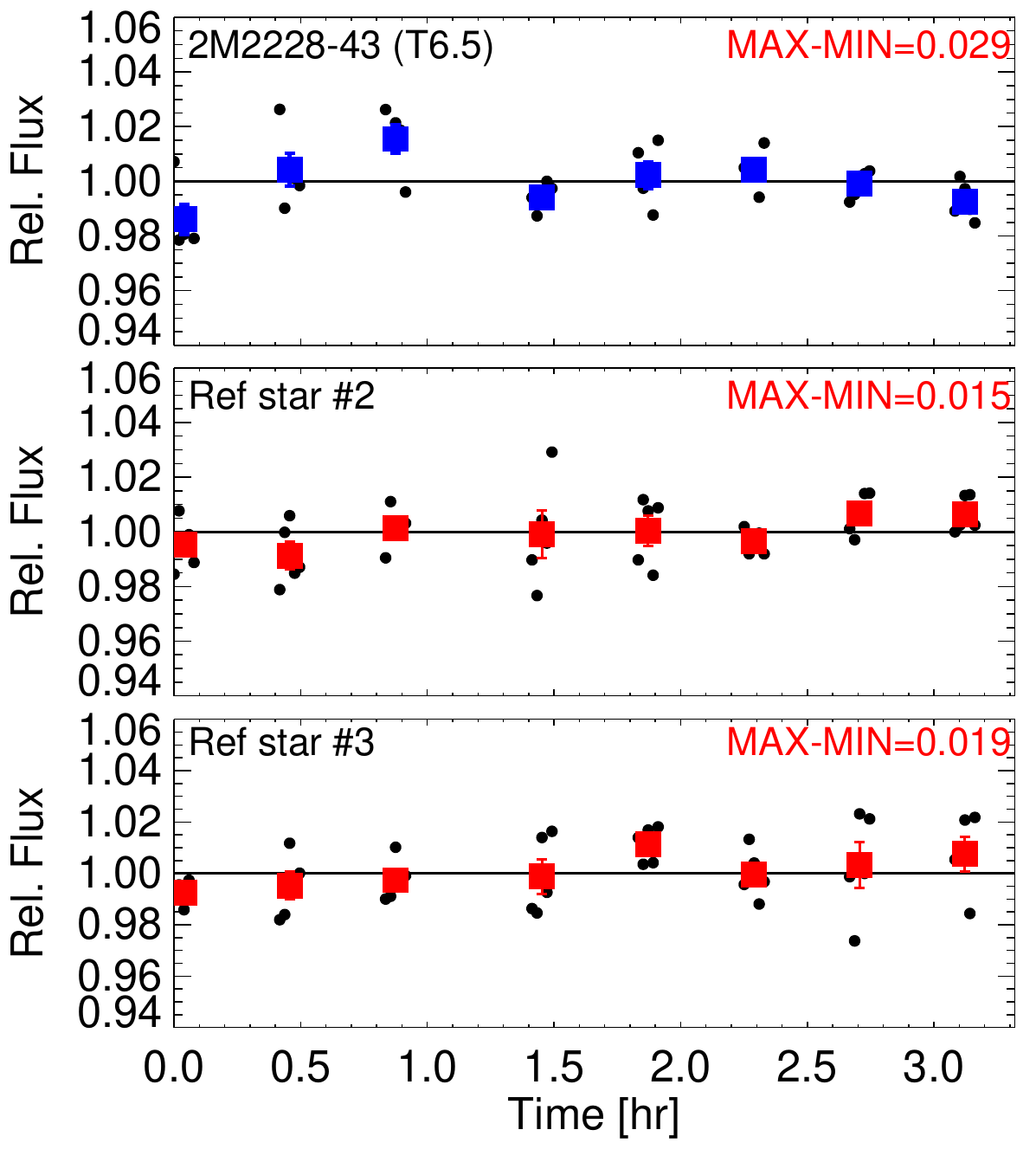}} &
\multirow{2}{*}[0in]{\hspace{-0.2 in} \includegraphics[width=0.3\hsize]{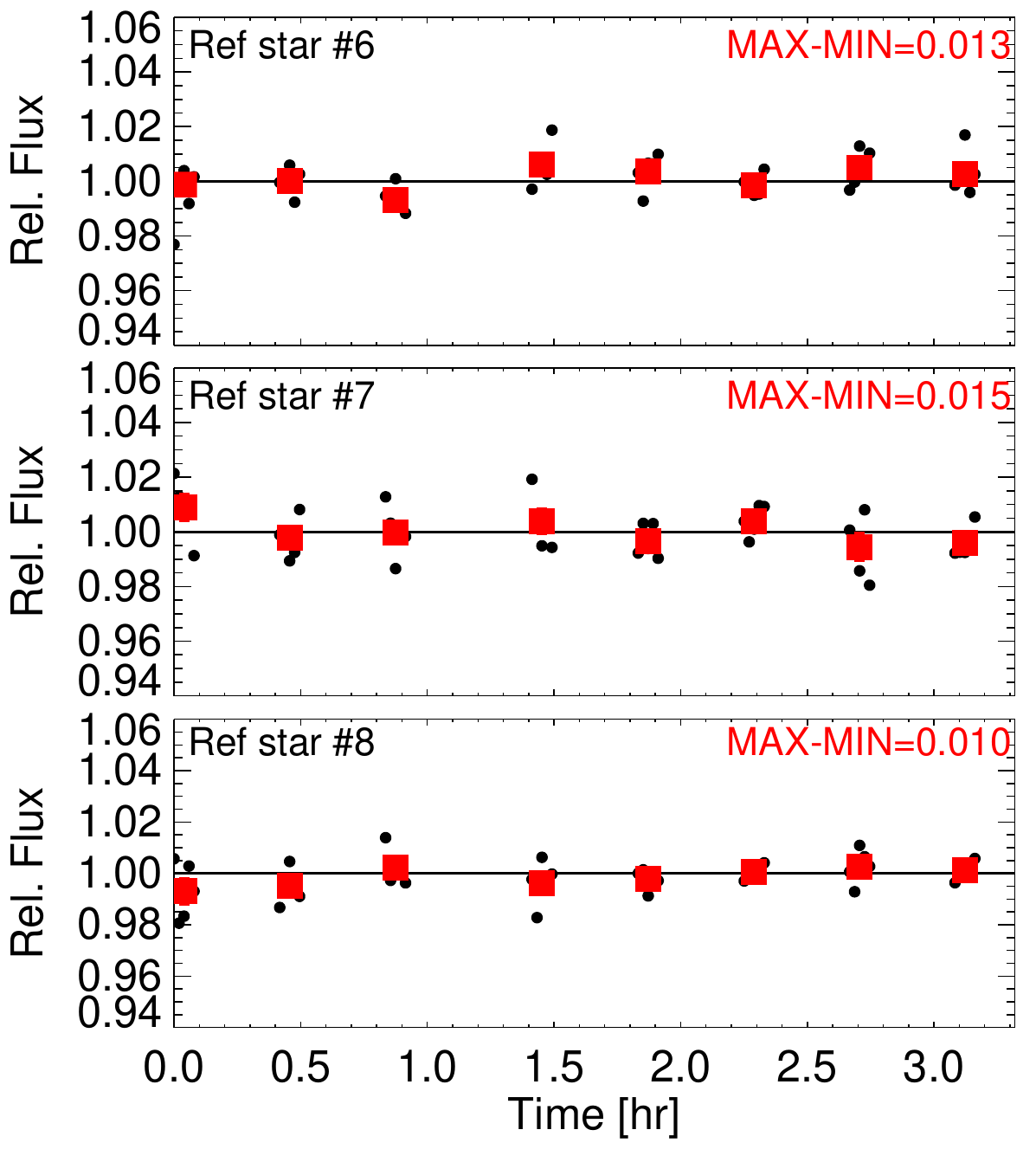}} &
\multirow{1}{*}[0.15in]{\hspace{-0.25in}\includegraphics[width=0.36\hsize]{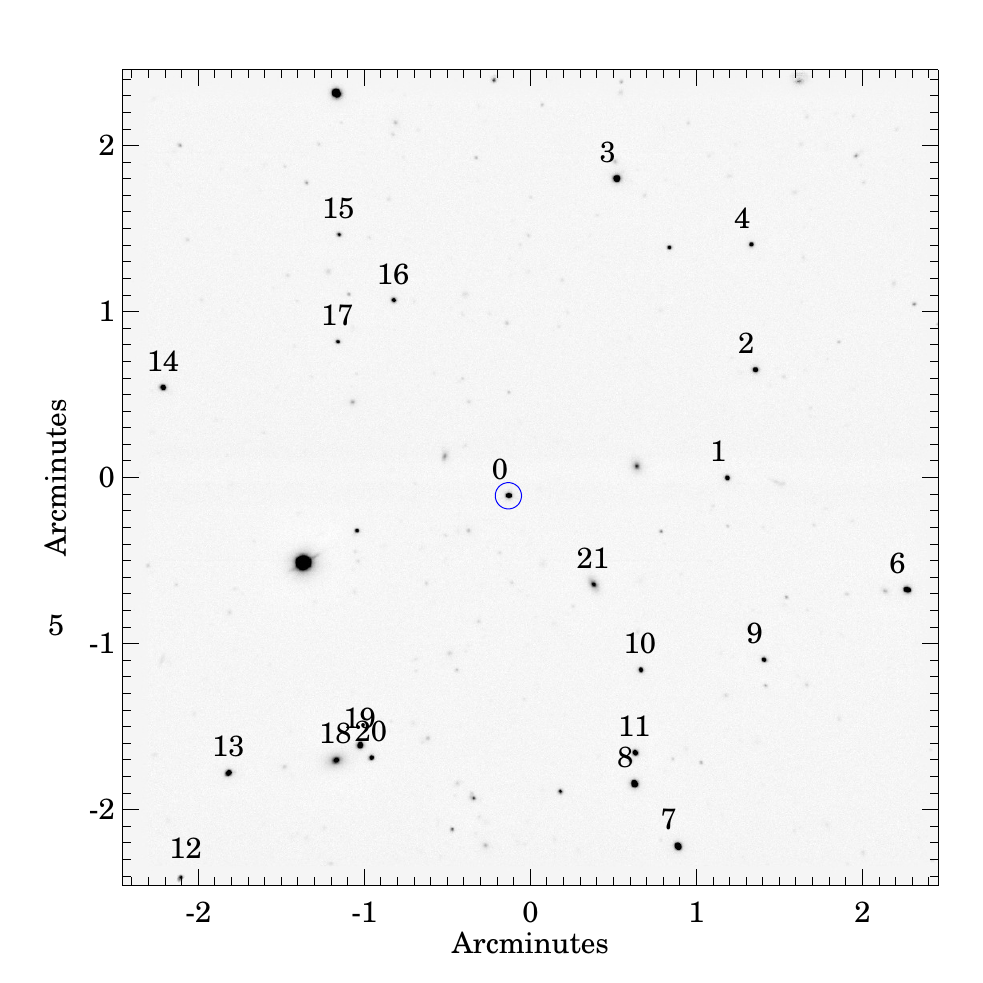}}
\end{tabular}
\vspace{2.3 in}\caption{Same as figure \ref{fig:lc1} but for the T6.5 dwarf 2M2228$-$43.  W14 report a target amplitude of 3.9$\pm$0.7\% for this time series.  \label{fig:lc13}}
\end{figure*}

\section*{Acknowledgements}
JR thanks Daniel Apai, {\'E}tienne Artigau, Jens Chluba, David Lafreni{\`e}re and Neill Reid for valuable discussion and feedback related to this work.   JR is supported by a Giacconi Fellowship through the Space Telescope Science Institute, which is operated by the Association of Universities for Research in Astronomy, Incorporated, under NASA contract NAS5-26555.  The data presented in this paper are based on observations made with ESO Telescopes at the La Silla Paranal Observatory under programme ID 188.C-0493. 

\bibliographystyle{/home/radigan/manuscripts/astronat/apj/apj}
\bibliography{mylib}

\end{document}